\documentstyle[preprint,eqsecnum,aps]{revtex}

\renewcommand{\baselinestretch}{1.0}
\begin{document}

\preprint{HD-TVP/9905}
\title{\vspace{1.5cm} Collision integral for multigluon production in a
scalar quark \& gluon model}
\author{D.~S. Isert and S.~P.~Klevansky}
\address{ Institut f\"ur Theoretische Physik, \\
Philosophenweg 19, D-69120 Heidelberg, Germany}

\maketitle
\vspace{1cm}

\begin{abstract}
A model of scalar gluons and scalar quarks that successfully gives a 
$\ln s$ behavior in high energy $q q$ scattering and which contains
a three gluon vertex is used to derive transport equations for the quarks
and gluons.   Quasiparticle and semiclassical approximations are utilized.
In particular, the collision integral is studied.  At first the terms are
organized according to the number of gluon exchange interactions, and it is shown 
explicitly that the occurrence of all two body differential scattering
cross sections in the Boltzmann equation requires an evaluation of 
{\it all} terms to this order.   One obtains $qg\rightarrow qg$, 
$q\bar{q} \rightarrow gg$, $qq
\rightarrow qq$, and $q\bar q\rightarrow q\bar q$.   An additional ordering
according to the inverse number of colors reveals that the quark loop
diagrams are suppressed, and diagrams that lead to gluon production or
scattering dominate.  
 The self energy with three gluons
exchanged is then examined.   From this, it is evident that initial as well
as
final state interactions enter into a generalized collision integral.  
The  generalized Boltzmann-like equation involving $n$ gluon exchange
for quark transport is given.

\end{abstract}
\vskip 0.3in
PACS Numbers : 12.40.-y; 05.20.Dd; 12.38.Mh; 24.85.+p.
\clearpage

\renewcommand{\baselinestretch}{1.3}
\footnotesize\normalsize

\section{Introduction}
\label{sec:intro}

A central issue in describing heavy ion collisions is the question of
particle fragmentation, for which there exist several models \cite{wong}.
From a theoretical point of view, a description of such a collision usually
is made within the framework of non-equilibrium theory in which the equation
governing the Wigner function that describes the particle distribution
 $f(X,p)$ semiclassically must somehow be given.   From our
knowledge of classical systems out of equilibrium and the attendant
Boltzmann equations, it is possible to make {\it heuristic ans\"atze\ }
for the equations governing the 
distribution functions of the relevant degrees of freedom.
This line of thought has been taken up by several authors in 
particular from the 
partonic point of view \cite{bmuller}, and, based on this, 
parton  cascade codes have been set up
that are  successful and necessary for the description of hard
processes.

On the other hand, a convincing formal justification of these heuristic
ideas
using Green function techniques does not yet exist, although steps have 
been taken in this direction in attempts to derive transport equations
directly from the underlying theory of the strong interactions, quantum
chromodynamics (QCD), see for example \cite{bmkg,klaus,heinz1,gyulassy}.
   By now, the basic equations of motion for Green
functions within a non-equilibrium approach \cite{bot,chou,landau}
 for any theory are well known:
they can simply be written down in a formal sense; one has a complicated set
of matrix Dyson equations relating matrices of Green functions with
 matrices of
self energies, for each particle species.   The crux of the issue, after
Wigner transformation, is the determination of these self energies for
a given theory.   The lowest order or mean field contributions to the 
self energy has been studied in several models of bosons or fermions or 
both, see for example \cite{ulli,perry}.
   Higher order terms that contribute to the self energy
and which can be interpreted as giving rise to collisional effects are less
well studied.    Recently, however, it was shown exactly within a simple
relativistic model of quarks, the Nambu--Jona-Lasinio  model \cite{nambu},
 that all
graphs that contribute to second order in the interaction are required to
generate two body elastic scattering diagrams that contribute to the
differential cross-sections occurring in the Boltzmann equation \cite{ogu}.

In this current paper, we study a partonic model of QCD
inspired by Polkinghorne \cite{polk} that contains scalar
quarks and scalar gluons that are coupled to each other.   In addition,
the gluons interact with each other via a triple gluon vertex, simulating
the non-abelian nature of QCD.   Such a 3-gluon vertex turns out to be
sufficient to produce ``pomeron-like'' behavior {\it i.e.} the quark-quark
elastic scattering amplitude can be calculated in a perturbative expansion
in leading logarithms of the center of mass energy $s$, and such an
expansion involves only gluonic ladder like diagrams constructed between
the two scattering quarks \cite{forsh}. 
    As it turns out, a 4-gluon
vertex would be subleading in the expansion in $\ln s$ and is therefore
omitted in this model Lagrangian.   We will take this as our starting point.
It has the advantage over QCD proper that we can study the consequences of
the non-abelian like terms, but we do not have to concern ourselves with the
additional problem of gauge invariance \cite{gyulassy,hz}.
   There is one additional 
difficulty, however, 
that the color structure of the theory becomes necessarily more
complicated, but this can be handled relatively easily.  

Using the scalar partonic model, we examine the self energies that are
required in the non-equilibrium formalism in and beyond the mean field
approximation for
use in constructing a collision term.  Within the mean field approximation,
the Hartree term is identically zero, due to color traces.
   The Fock term, which is often
ignored in such calculations\footnote{See however Refs\cite{henning}
and \cite{rehberg}}, is discussed in some detail.   The assumption 
that this term vanishes also, leads to enormous simplifications in considering
higher order terms in the coupling $g$.  We then
examine the two gluon exchange diagrams, of which there are several
different kinds.   We classify them according to their topology as being
of the ``rainbow'', ``cloud'', ``ladder'', ``exchange'' or ``quark loop'' kind.
We find 
  that {\it all} the graphs occurring
in this order are required in order to produce the two body partonic
differential scattering cross sections that would enter into a 
Boltzmann-like transport equation, in contrast to that suggested in
\cite{klaus}.  This is an exact result.   In our case, regarding for
example the equations of motion for the quark distribution function,
we find that the relevant 
cross sections that enter into the collision term at this level 
are either purely quark like in nature ($q\bar q\rightarrow
q\bar q$ and $qq\rightarrow qq$), or contain gluon components
($qg\rightarrow qg$ and $q\bar q \rightarrow gg$)\footnote{Other processes,
such as $q\bar q g\to g$ and
$q\bar q gg\to 0$ are suppressed by kinematical considerations.}.   
Due to the plethora of different types of graphs that already occur at the
two gluon exchange level, we seek an additional expansion parameter
to simplify this and to enable us to go to higher orders in the coupling.
In QCD itself, an expansion in the inverse number of colors is often
useful, and we thus examine such an expansion within this model.
   An analysis of the color structure of each graph
naturally suppresses diagrams containing quark loops and the exchange graphs
thereof, as is also a feature of QCD.   All remaining
  self energy graphs that  contain gluonic exchange components (the ladder,
rainbow and cloud graphs) turn out to be
 superficially of the same order in $N_c$,
and all are required to generate all possible types of gluon production
or quark-gluon scattering.   
A closer look, however, shows that the ratio of the ladder to the rainbow or 
cloud graphs goes as the ratio of the Casimir operators of the adjoint to the 
fundamental representations squared,
  which favors the ladder graphs numerically.
The two body process is then briefly reanalysed along these lines.

With this point of view, the next order in the coupling, i.e. the three
 gluon exchange is investigated.  
Examining once again the equation of motion for the quark distribution
function, we find that three parton production, {\it i.e.} $q\bar q\to g g
g$ and $qg\to qgg$, naturally occurs.   In addition to these production
processes, however, the processes $q\bar q\to gg$ and $qgg\to qg$ must
also be accounted for\footnote{Again, such processes
 as $q\bar q ggg\to 0$ and
$q\bar q gg\to g$ are suppressed by kinematical considerations.}.
We thus see that initial as well as final
state processes must be included in the matrix elements. 
 With this information,  the generalized Boltzmann equation that includes 
multiparton production ($q\bar q\to ng$, but also processes of the 
form $q\bar q mg\to ng$, with $m$ and $n$ integer) 
within this scalar quark and gluon model is given.

   It is to be noted that, unlike the situation that occurs in evaluating
the elastic quark-quark scattering amplitude within the scalar
parton model \cite{forsh}, we cannot {\it a priori} make 
an assumption on the internal ordering of momenta within a self energy graph.
This comes about in the elastic quark-quark scattering amplitude, because
there are two external momenta that are fixed.  In a self energy graph,
however, and in the Boltzmann equation, 
only one external momentum can be fixed while {\it all}
the remaining momenta are integrated over.  Thus the simplifications leading
to the ladder approximation and Regge behavior in that calculation
 are not simply
recovered in the kinetic theory. 

This paper is organized as follows.   In Section \ref{sec:model}, we give a
brief introduction to the scalar partonic model, and set our notation for
the nonequilibrium dynamics.    In Section \ref{sec:mean}, we describe the
collisional contributions that arize from the
mean field self energies, and the associated transport equations.   In Section
\ref{sec:collision}, we deal with the collision integral beyond the mean
field approach, firstly giving the
exact perturbative calculations with two gluon exchange, the color
expansion, then three and $n$ gluon exchange.   We summarize and conclude 
in Section \ref{sec:summary}.

\section{Model of scalar quarks and gluons, and transport theory}
\label{sec:model}

In this section, we introduce and discuss the scalar partonic model, and
give the equations of motion for quark and gluon fields.   We then give
a brief discussion of transport theory applied to this scalar model, and
indicate the self energies that we are required to evaluate.

\subsection{Scalar partonic model}

In the model underpinning our calculations, quarks and antiquarks are
described by complex scalar fields
 $\phi$, and gluons as the  scalar field $\chi$ coupled through the Lagrangian
\begin{eqnarray}
{\cal L}&=& \partial^{\mu}{\phi}^{\dagger i,l}\partial_{\mu}\phi_{i,l} +
\frac{1}{2}\partial^{\mu}\chi_{a,r}\partial_{\mu}\chi^{a,r}
-\frac{m^2}{2}\chi_{a,r}\chi^{a,r} \nonumber \\
&& - g m{\phi}^{\dagger i,l}(T^a)^j_i(T^r)^m_l\phi_{j,m}\chi_{a,r} -
\frac{gm}{3!} f_{abc}f_{rst} \chi^{a,r}\chi^{b,s}\chi^{c,t}.
\label{e:lagrangian}
\end{eqnarray}

The quark fields are regarded as  massless,
as one generally assumes for high energy processes, 
while the gluons are usually assigned  a mass $m$ {\it a priori}
in order to avoid infra-red divergences. 
There is an interaction between quarks and gluons as well as a
cubic self-interaction between gluons. Since in QCD the quartic interaction
between gluons leads to terms which are sub-leading in ln $s$, such a
quartic interaction among gluons is not included within this model. 

One notes that both the quark and gluon fields carry two labels.   Both of
these refer to color groups.   The fact that a direct product of two color 
groups is necessary can be seen on examining the three gluon vertex term.
 This  term must be symmetric under the exchange of two gluons since they 
are bosons.  In addition, one expects that the
interaction vertex should be proportional to the (antisymmetric) structure 
constants of the color group.  A single color group cannot meet these 
requirements, and the simplest combination which can 
is a product of two $SU(N_c)$ groups. Thus the gluon field
carries two color indices ($a, r = 1...(N_c^2 - 1)$). Since the quark field
transforms in the fundamental representation of both of these $SU(N_c)$
groups,
it must carry two color indices as well ($i,l= 1...N_c$).
The matrices $T^a$ and $T^r$ are the generators and $f_{abc}$ and $f_{rst}$
are the structure constants for the two $SU(N)$ groups respectively.
Thus they satisfy
\begin{equation}
[T^a,T^b]=if_{abc}T^c,\quad\quad [T^r,T^s]=if_{rst}T^t.
\label{e:commutator}
\end{equation}
Note in Eq.(\ref{e:lagrangian}) that the flavor index of the quark fields
is suppressed.
    
The equations of motion for the fields can be derived from the Euler-Lagrange
equations. They are 
\begin{equation}
\Box \phi^{(\dagger)i,l} = -gm(T^a)^i_j(T^r)^l_m \phi^{(\dagger)j,m} \chi_{a,r}
\label{motionq}
\end{equation}
for the (anti-)quarks and  
\begin{equation}
(\Box + m^2)\chi^{a,r} = -gm[\phi^{\dagger i,l}(T^a)^i_j(T^r)^m_l\phi_{j,m}
                              + f^{abc}f^{rst}\chi_{b,s}\chi_{c,t}]
\label{motiong}
\end{equation}
for the gluons.

\subsection{Transport theory for interacting bosonic fields}

For the purpose of establishing our notation, we give the basic
definitions and refer the reader to standard texts \cite{bot,chou,landau}.
   The quark Green functions in the Schwinger-Keldysh formalism
\cite{schwing}
are defined as
\begin{eqnarray}
iS^c(x,y)=&\left \langle T\phi^{i,l}(x)\phi^{\dagger j,m}(y)\right \rangle
-\left \langle \phi^{i,l}(x)\right \rangle
 \left \langle \phi^{\dagger j,m}(y)\right \rangle&=iS^{--}(x,y)\nonumber\\
iS^a(x,y)=&\left \langle \tilde{T}\phi^{i,l}(x)\phi^{\dagger j,m}(y)\right
\rangle
-\left \langle \phi^{i,l}(x)\right \rangle
 \left \langle \phi^{\dagger j,m}(y)\right \rangle&=iS^{++}(x,y)\nonumber\\
iS^>(x,y)=&\left \langle \phi^{i,l}(x)\phi^{\dagger j,m}(y)\right \rangle
-\left \langle \phi^{i,l}(x)\right \rangle
 \left \langle \phi^{\dagger j,m}(y)\right \rangle&=iS^{+-}(x,y)\nonumber\\
iS^<(x,y)=&\left \langle \phi^{\dagger j,m}(y)\phi^{i,l}(x)\right \rangle
-\left \langle \phi^{i,l}(x)\right \rangle
 \left \langle\phi^{\dagger j,m}(y)\right \rangle&=iS^{-+}(x,y)
\label{e:defs}
\end{eqnarray}
and for the gluons as
\begin{eqnarray}
iG^c(x,y)=&\left \langle T\chi^{a,r}(x)\chi^{b,s}(y)\right \rangle
-\left \langle \chi^{a,r}(x)\right \rangle
 \left \langle \chi^{b,s}(y)\right \rangle
&=iG^{--}(x,y)\nonumber\\
iG^a(x,y)=&\left \langle \tilde{T}\chi^{a,r}(x)\chi^{b,s}(y)\right \rangle
-\left \langle \chi^{a,r}(x)\right \rangle
 \left \langle \chi^{b,s}(y)\right \rangle &=iG^{++}(x,y)\nonumber\\
iG^>(x,y)=&\left \langle \chi^{a,r}(x)\chi^{b,s}(y)\right \rangle
-\left \langle \chi^{a,r}(x)\right \rangle
 \left \langle \chi^{b,s}(y)\right \rangle
&=iG^{+-}(x,y)\nonumber\\
iG^<(x,y)=&\left \langle \chi^{b,s}(y)\chi^{a,r}(x)\right \rangle
-\left \langle \chi^{a,r}(x)\right \rangle
\left \langle \chi^{b,s}(y)\right \rangle
&=iG^{-+}(x,y).
\label{e:defg}
\end{eqnarray}
Here $T$ and $\tilde{T}$ are the usual time and anti-time ordering operators
respectively.   As given, the Green functions fall along the contour
designated in Fig.~1.   Our convention follows that of
Ref.\cite{landau}.
   They satisfy a Dyson equation that introduces the
matrix of self energies for either the quark or gluonic sectors,
$\underline \Sigma_q$ or $\underline \Sigma_g$.  Using a generic notation,
$\underline{D}
 = \underline S$ or $\underline G$ and ${\underline{\Pi}}= {\underline \Sigma_q}$ or
 $\underline \Sigma_g$ as appropriate, one may write 
\begin{eqnarray}
{\underline D}(x,y) &=& {\underline D}^0(x,y) - \int d^4z d^4w {\underline
D^0(x,w)}{\underline \Pi}(w,z){\underline D}(z,y) \nonumber \\
 &=& {\underline D}^0(x,y) - \int d^4z d^4w {\underline
D}(x,w){\underline \Pi}(w,z){\underline D}^0(z,y). \nonumber \\
\label{e:dyson}
\end{eqnarray}
   In a standard fashion, the
transport and constraint equations can be derived, and this is summarized
briefly in Appendix A.   In terms of the center of mass variable
$X=x+y$ and the Fourier transform variable $p$ conjugate to the relative
distance $u=x-y$, or Wigner transform, 
 the equations that one obtains for the off diagonal
Green functions, corresponding to the transport and constraint equations
are
\begin{equation}
-2ip\partial_X D^{-+}(X,p)=I_- \quad\quad {\rm transport}
\label{transport}
\end{equation}
and
\begin{equation}
\left( \frac{1}{2} \Box_X - 2p^2 + 2M^2\right) D^{-+}(X,p) = I_+,  \quad\quad
{\rm constraint}
\label{constraint}
\end{equation}
respectively,
where $M$ is a generic parton mass, $M=m$ for the gluons and $M=0$ for the
quarks.  $I_{\mp}$ is an abbreviation for the combined functions
\begin{equation}
I_{\mp}=I_{\rm coll} + I_{\mp}^A + I_{\mp}^R,
\label{e:imp}
\end{equation}
and $I_{\rm coll}$ is the collision term,
\begin{eqnarray}
I_{\rm coll} &=& \Pi^{-+}(X,p)\hat{\Lambda}D^{+-}(X,p)
                 -\Pi^{+-}(X,p)\hat{\Lambda}D^{-+}(X,p)\nonumber\\
             &=& I_{\rm coll}^{\rm gain} - I_{\rm coll}^{\rm loss}.
\label{e:collision}
\end{eqnarray}
$I_{\mp}^R$ and $I_{\mp}^A$ are  terms containing retarded and 
advanced components respectively,
\begin{equation}
I_{\mp}^R = -\Pi^{-+}(X,p)\hat{\Lambda}D^{R}(X,p)
            \pm    D^{R}(X,p)\hat{\Lambda}\Pi^{-+}(X,p)
\label{e:rim}
\end{equation}
and
\begin{equation}
I_{\mp}^A = \Pi^{A}(X,p)\hat{\Lambda}D^{-+}(X,p)
            \mp    D^{-+}(X,p)\hat{\Lambda}\Pi^{A}(X,p).
\label{e:pim}
\end{equation}
In Eqs.(\ref{e:collision}) to (\ref{e:pim}), the operator $\hat \Lambda$ is
given by
\begin{equation}
\hat{\Lambda}:={\rm exp}\left\{ \frac{-i}{2}\left(\overleftarrow{\partial}_X
 \overrightarrow{\partial}_p
-\overleftarrow{\partial}_p\overrightarrow{\partial}_X\right) \right\}.
\label{lambda1}
\end{equation}         
In order to proceed further, we have to calculate the self energies.
It is useful to introduce the quasiparticle approximation, in which a
{\it free} scalar parton of mass $M$ is assigned the Green functions
\begin{eqnarray}
iD^{-+}(X,p)&=&\frac{\pi}{E_p}\{\delta(E_p-p^0)f_a(X,p)+\delta(E_p+p^0)
\bar{f}_{\bar a} (X,-p)\}\label{d-+}\\                                     
iD^{+-}(X,p)&=&\frac{\pi}{E_p}\{\delta(E_p-p^0)\bar{f}_a(X,p)+\delta(E_p+p^0)f
_{\bar a}(X,-p)\}\label{d+-}\\
iD^{--}(X,p)&=&\frac{i}{p^2-M^2+i\epsilon}+\Theta(-p^0)D^{+-}(X,p)+
\Theta(p^0)D^{-+}(X,p)\nonumber\\
&=&\frac{i}{p^2-M^2+i\epsilon}+\frac{\pi}{E_p}\{\delta(E_p-p^0)f_a(X,p)+
\delta(E_p+p^0)f_{\bar a}(X,-p)\}\nonumber \\ \label{d--}\\
iD^{++}(X,p)&=&\frac{-i}{p^2-M^2-i\epsilon}+\Theta(-p^0)D^{+-}(X,p)+
\Theta(p^0)D^{-+}(X,p)\nonumber\\               
&=&\frac{-i}{p^2-M^2-i\epsilon}+\frac{\pi}{E_p}\{\delta(E_p-p^0)f_a(X,p)+
\delta(E_p+p^0)f_{\bar a}(X,-p)\}\nonumber \\ \label{d++}
\end{eqnarray}
with $E_p^2=p^2+M^2$, and 
which are given in terms of the corresponding scalar quark and gluon 
distribution function,
$f_a(X,p)$, and $\bar f_a = 1+f_a$, where $a$ denotes the parton type $a=q,g$.

Our task in this paper is 
to construct an equation for the distribution functions for
quarks and gluons $f_a(X,p)$ from Eqs.(\ref{transport}) to (\ref{e:pim}),
using the quasiparticle Green functions of the form given in Eqs.(\ref{d-+})
to (\ref{d++}).  To do so, it is necessary to integrate  
the entire Eq.(\ref{transport}) over an interval $\Delta_\pm$ which
contains $\pm E_p(X)$.  
Our particular focus will be placed on the collision terms that are derived
from Eq.(\ref{e:collision}).
 In anticipation of this fact, we will require such an integration over the
terms appearing in Eq.(\ref{e:collision}), and define
\begin{eqnarray}
J_{\rm coll} & = & J^{\rm gain}_{\rm coll}-J^{\rm loss}_{\rm
coll}\nonumber\\  
&=& \int_{\Delta^+}\! dp_0\,I^{\rm gain}_{\rm coll}
   -\int_{\Delta^+}\! dp_0\,I^{\rm loss}_{\rm coll}.
\label{e:j22}
\end{eqnarray}
To lowest order in an expansion that sets $\hat \Lambda = 1$, this integral
can be easily performed.   One has
\begin{eqnarray}
J_{\rm coll}
&=& \int_{\Delta^+}\! dp_0\,\Pi^{-+}(X,p)\,D^{+-}(X,p)
   -\int_{\Delta^+}\! dp_0\,\Pi^{+-}(X,p)\,D^{-+}(X,p)\nonumber\\
&=& -i
\frac{\pi}{E_p}\,\Pi^{-+}(X,p_0=E_p,\vec{p})\,\bar{f}_a(X,\vec{p})
    +i \frac{\pi}{E_p}\,\Pi^{+-}(X,p_0=E_p,\vec{p})\,f_a(X,\vec{p}),
\nonumber\\
\label{e:onshell2}
\end{eqnarray}
{\it i.e.} the off-diagonal quasiparticle self energies are required to be
calculated on shell.

\section{The Collision Integral - Mean field self energies}
\label{sec:mean}

In this section, we evaluate the self energies to first order
in the interaction strength and illustrate their role in the transport
equation in the semi-classical limit.  

\subsection{Hartree self energies}

For the scalar parton model, two generic kinds of Hartree 
graphs can be identified
in the quark and gluon self energies.  These are depicted in
Fig.~2. 
For Hartree diagrams of any kind, off diagonal self energies are
per definition zero and  only diagonal elements can possibly be 
constructed, {\it i.e.} $\Sigma^{--}_H$ or $\Sigma^{++}_H$.
 However, all such diagrams vanish identically in this model.    The reason
for this lies in the color factors:   for the quark self energy graph in
Fig.~2 that contains a quark loop, a single SU(3) color group
leads to the associated color factor
\begin{equation}
F_{H,q}= t^a_{ii}{\rm tr} (t^a)= 0,
\end{equation}          
since $t_a=\lambda^a/2$, where $\lambda^a$ are the Gell-Mann matrices.
In the above expression, $i$ denotes the external quark momentum and is
therefore not to be summed over.   For the quark self energy containing a
gluon line, the color factor for a single SU(3) group is also vanishing,
\begin{equation}                 
F_{H,g}= t^a_{ii} T^a_{bb} = -i t^a_{ii} f_{abb} =0 .
\end{equation}          
Similar arguments apply to the gluon self energies.   Thus, no mass
renormalization occurs due to Hartree terms.

A semiclassical expansion of the transport and constraint equations thus 
leads to  free streaming described by
\begin{equation}
-2i p\partial_X D^{-+}(X,p) = 0,
\label{e:free}
\end{equation}
and the constraint equation is
\begin{equation}
(p^2-M^2) D^{-+}(X,p) =0.
\label{e:freecon}
\end{equation}
By integrating Eq.(\ref{e:free}) over a positive interval $\Delta_+$, the 
 equations for the time evolution of the
distribution function for quarks and gluons emerge,
\begin{equation}
p\partial_Xf_{a}(X,p) =0, 
\end{equation}
with 
$p^\mu = (E_p,\vec p)$.    On integrating the constraint equation,
Eq.(\ref{e:freecon}), one has $(E_p^2 - \vec p^2-M^2)f_{a}(X,p)=0$.
The  equation is the expression of the fact 
 that the partons have to be on mass-shell,
and is consistent with the quasiparticle assumption, Eqs.(\ref{d-+}) -
(\ref{d++}) made in the first place.

\subsection{Fock self energies}

The next type of graph contributing to the mean field  expansion is the Fock
 term.
The generic diagrams for the quark and gluon self energies are shown in
 Fig.~3\footnote{Fig.3(b) is strictly speaking a vacuum
polarization graph for the gluons}.

Since we are particularly concerned with the collision integral, we start
with the quark sector and examine as an example, the gain term generated
by the Fock term $\Sigma^{-+}_{F,q}(X,p)$.   By inspection, one has
\begin{eqnarray}
i\Sigma_{F,q}^{-+}(X,p) &=&-g^2m^2 F_{F,q}^2\int\!
\frac{d^4p_1}{(2\pi)^4}\int\!\frac{d^4p_2}{(2\pi)^4} 
S^{-+}(X,p_1) G^{+-}(X,p_2)(2\pi)^4\delta^{(4)}(p\!-\!p_1\!+\!p_2),\nonumber \\
\label{e:sigf}
\end{eqnarray}
and $F_{F,q}$ is the Fock color factor for a single SU($N_c$) group,
\begin{equation}
F_{F,q} = t^a_{ij}t^a_{ji} = \frac{N_c^2-1}{2N_c}\delta_{ii}.
\end{equation}
The contribution to the collision term that this makes, using
Eqs.(\ref{e:j22}) and (\ref{e:onshell2}) is
\begin{equation}
J^{{\rm gain}}_{{\rm F, coll}}= \int_{\Delta^+} dp^0 \Sigma^{-+}_{{\rm F,q}}
(X,p) (-\frac {i\pi}{E_p})\delta(p_0-E_p)\bar f_q(X,p),
\label{e:good}
\end{equation}
which, on inserting the explicit expressions for $S^{-+}(X,p)$ and
$G^{+-}(X,p)$
from
Eqs.(\ref{d-+}) and (\ref{d+-}) leads to four distinct terms,
\begin{eqnarray}
J^{{\rm gain}}_{{\rm F, coll}} &=&- g^2m^2F_F^2\frac{\pi}{E_p}\int_{\Delta^+} 
dp^0 \delta(E_p^q -p^0) \int                                           
\frac{d^4p_1}{(2\pi)^4}\int\frac{d^4p_2}{(2\pi)^4}
(2\pi)^4\delta^{(4)}(p+p_1-p_2) 
\nonumber \\ 
&\times&\frac{\pi}{E_{1}} \frac{\pi}{E_{2}} \sum_{i=1}^4 T_i,
\label{e:fourterms}
\end{eqnarray}
where
\begin{eqnarray}
T_1 &=& \delta(E_{1}-p_1^0)\delta(E_{2}-p_2^0) \bar f_{\bar q}
(X,p_1)f_g(X,p_2)
\bar f_q(X,p) \nonumber \\
T_2 &=& \delta(E_{1}-p_1^0)\delta(E_{2}+p_2^0) \bar f_{\bar q}(X,p_1)\bar
f_g(X,-p_2) \bar f_q(X,p) \nonumber \\
T_3 &=& \delta(E_{1}+p_1^0)\delta(E_{2}+p_2^0) f_q(X,-p_1) \bar
f_g(X,-p_2) \bar f_q(X,p) \nonumber \\
T_4 &=& \delta(E_{1}+p_1^0)\delta(E_{2}-p_2^0) f_q(X,-p_1) f_g(X,p_2) \bar
f_q(X,p).
\label{e:tfocks}
\end{eqnarray}
By attributing unbarred functions $f$ to incoming particles and barred
functions $\bar f$ to outgoing ones, one can see that $T_1..T_4$ correspond
to the processes $g\rightarrow q\bar q$, {\O}$\rightarrow q\bar q g$,
$q\rightarrow qg$ and $qg\rightarrow q$.  The last three of these are 
kinematically forbidden, while the former is possible, since the gluons are
endowed with a finite mass.
Performing the integrals over $p_1^0$ and $p_2^0$, Eq.(\ref{e:tfocks}) 
becomes
\begin{eqnarray}
\int_{\Delta^+} dp^0 \Sigma_{F,q}^{-+}(X,p) S^{+-}(X,p)& =&
-\frac{\pi}{E_p} \int\frac{d^3p_1}{(2\pi)^3 2E_1} \frac {d^3p_2}{(2\pi)^3
2E_2} (2\pi)^4 \delta^{(4)}(p+p_1-p_2)\nonumber \\
&& \times |{\cal M}|^2_{g\to q\bar q}
f_g(X,p_2) \bar f_{\bar q}(X,p_1) \bar f_q(X,p).
\label{e:fintegrated}
\end{eqnarray}
The loss term is obtained in a similar fashion or by exchanging $f$ with
$\bar f$, since the matrix element is symmetric.    Combining both terms,
the revised transport equation is obtained on integrating
Eq.(\ref{e:collision}) over the energy interval $\Delta_+$ as
\begin{eqnarray}
2p\partial_X f_q(X,p)& =& \int
\frac{d^3p_1}{(2\pi)^32E_1}\frac{d^3p_2}{(2\pi)^32E_2} (2\pi)^4\delta^{(4)}
(p+p_1-p_2)\nonumber \\
&&\times |{\cal M}|^2_{g\rightarrow q\bar q} [ f_g(X,p_2)\bar f_{\bar q}
(X,p_1) \bar f_q(X,p) - \bar f_g(X,p_2) f_q(X,p_1)  f_q(X,p)].
\nonumber \\
\label{e:fockcoll}
\end{eqnarray}
This is the final expression for the Fock transport equation.  One can 
alternatively introduce a cosmetic recombination or decay rate in which case
Eq.(\ref{e:fockcoll}) can be written symbolically as
\begin{eqnarray}
2p\partial_X f_q(X,p) &=& \int\frac{d^3p_2}{(2\pi)^32E_2}\int d\Omega
\frac{d\sigma}{d\Omega}\bigg|_{q\bar q\to g} F \nonumber \\
&& \times[
 f_g(X,p_2) \bar f_{\bar q}(X,p_1) \bar f_q(X,p) - \bar f_g(X,p_2)
f_q(X,p_1)  f_q(X,p)],
\label{e:focktrans}
\end{eqnarray}
where $F$ is the flux factor, and  
\begin{equation}
\int d\Omega \frac {d\sigma}{d\Omega} = \int dQ \frac{|{\cal M}|^2}{F}
\label{e:om}
\end{equation}
with the invariant phase space factor $dQ$ given as
$dQ=(2\pi)^4\delta^{(4)}(p+p_1-p_2)\, d^3p_2/((2\pi)^32E_2)$.

An analysis of the self energy graph 3(a)of the gluon sector,
$\Sigma^{-+}_{F,g (a)}(X,p)$ along the previous lines leads to processes
$g\rightarrow gg$, {\O}$\rightarrow ggg$, and $gg\rightarrow g$, all of
which are kinematically prohibited.   One thus obtains
\begin{equation}
(J^{{\rm coll} (a)}_{{\rm F, gain/loss}})_{{\rm gluonic \  graph}} = 0.
\end{equation}
This can be attributed to the fact that that the self energies are
evaluated on shell, i.e. we may write
\begin{equation}
\Sigma^{-+}_{F,g(a)}(X,p_0=E_p,\vec p) = 
\Sigma^{+-}_{F,g(a)}(X,p_0=E_p,\vec p) =0,
\label{e:foff}
\end{equation}
which is the statement that an on shell particle cannot decay into two
on shell particles of the same kind.

The second graph in the gluonic case does not vanish.   This self energy
$\Sigma^{-+}_{F,g(b)}$ that enters into the description of the gain in
gluons, is precisely that given in Eq.(\ref{e:sigf}), but with $G^{+-}
(X,p_2)$ replaced by $S^{+-}(X,p_2)$.   The color factor in this case
is also modified, being $F_{F,g} = 1/2 \delta^{aa}$.
    An analysis of the self energy along
the same lines leads to the processes $q\to qg$, $\bar q \to \bar q g$,
{\O}$\to gq\bar q$ and $q\bar q\to g$, the last of which is the only term
that can contribute.   Thus the time evolution of the gluon distribution 
function is given by
\begin{eqnarray}
2p\partial_X f_g(X,\vec p)& =& \int
\frac{d^3p_1}{(2\pi)^32E_1}\frac{d^3p_2}{(2\pi)^32E_2} (2\pi)^4\delta^{(4)}
(p+p_1-p_2)\nonumber \\
&&\times |{\cal M}|^2_{g\rightarrow q\bar q} [ f_q(X,p_2) f_{\bar q}
(X,p_1) \bar f_g(X,p) - \bar f_q(X,p_2) \bar f_{\bar q}(X,p_1) f_g(X,p)].
\nonumber \\
\label{e:fockcoll2}
\end{eqnarray}

We conclude this section by commenting the result that while the
Fock term 3(a) for gluons vanishes identically, the Fock term for the quark
self energy does not.   A term of this kind occurs in this model because
the quarks are massless, while the gluons are massive.    The relevance of
this Fock term thus depends on the form of the underlying theory.   In 
evaluating higher order graphs in the following section, it is useful to
invoke the so-called ``on-shell argument'', ignoring all terms of higher
order of this kind 
that contain on-shell decays.   This is necessary in order to render
the problem tractable, but it is not strictly true, as has been indicated
here.

\section{The Collision Integral - beyond the mean field}
\label{sec:collision}
 
\subsection{Exact perturbative calculations that contain two loops
\label{sec4a}}

Generic diagrams that contain two loops and which contribute to
the quark and gluon sectors are shown in Fig.~4.   
In the quark sector,  we  denote these graphs as rainbow (R), 
ladder (L), cloud (C), exchange (E) and quark  loop graphs (q-Loop),
according to their topology.   These graphs comprise all diagrams with two
exchanged gluons.   In the gluonic sector, there are more diagrams, as
can be seen in the figure.  Note that in this sector, a ladder-like 
diagram is topologically equivalent to the rainbow kind, and is therefore
not included separately.
Since we are specifically interested in the collisional term, we will
concern ourselves only with the off-diagonal contributions that arise from
these diagrams, as is required in Eq.(\ref{e:collision}).
In addition, these self energies are required on shell.

To be specific, let us deal first with
the quark sector, and consider the rainbow diagram first.
   Then there are four possibilities of ordering the
lines in this graph, as is indicated in Fig.~5.    However,
invoking the on-shell argument,
 as explained in the previous section for the 
Fock diagram,
 only the first graph of this series can  survive while
 the other three vanish.   Similar arguments can be applied to the 
remaining graphs, the L, C, E, and q-Loop diagrams.    One can set
up all possibilities of ordering the lines for each type of graph
in Fig.~4, and
determine which survive.   The final set of all non-vanishing contributions
can then be ascertained.  
These are shown in Fig.~6.

We next note that the
 diagram of Fig.~6.4(b) is the exchange graph of 
Fig.~6.5, because both of these diagrams
 contain three off-diagonal quark Green
functions. The similarity becomes more evident if one redraws the diagram of
Fig.~6.4(b) as shown in Fig.~7.

The self energies for two gluon exchange contribute to the quark collision
term via
\begin{eqnarray}
J^{(2)}_{\rm coll}
&=& \int_{\Delta^+}\! dp_0\,\Sigma^{(2)-+}(X,p)\,S^{+-}(X,p)
   -\int_{\Delta^+}\! dp_0\,\Sigma^{(2)+-}(X,p)\,S^{-+}(X,p)\nonumber\\
&=& -i \frac{\pi}{E_p}\,\Sigma^{(2)-+}(X,p_0=E_p,\vec{p})\,\bar{f}_q(X,\vec{p})
    +i \frac{\pi}{E_p}\,\Sigma^{(2)+-}(X,p_0=E_p,\vec{p})\,f_q(X,\vec{p}),
\nonumber\\
\label{e:onshell}
\end{eqnarray}
where the off-diagonal self energies are constructed from the diagrams
of Fig.~4 as
\begin{equation}
\Sigma^{(2)\pm\mp}(X,p) = \Sigma^{(2)\pm\mp}_{{\rm quark-quark}}(X,p)
 + \Sigma^{(2)\pm\mp}_{{\rm quark-gluon}}(X,p)
\label{e:sumsig}
\end{equation}
where 
\begin{equation}
\Sigma^{(2)+-}_{{\rm quark-quark}}(X,p) =
\Sigma^{(2)+-}_{E,b)}(X,p) + \Sigma^{(2)+-}_{{\rm q-Loop}}(X,p)
\label{e:first}
\end{equation}
and 
\begin{eqnarray}
\Sigma^{(2)+-}_{{\rm quark-gluon }}(X,p) &=& \Sigma^{(2)+-}_R(X,p) +
\Sigma^{(2)+-}_L(X,p) + \Sigma^{(2)+-}_{C,a)}(X,p) \nonumber \\
&+& \Sigma^{(2)+-}_{C,b)}(X,p) +  \Sigma^{(2)+-}_{E,a)}(X,p).
\label{e:second}
\end{eqnarray}
Here we have split up the combinations of the second exchange plus quark
loop diagrams in $\Sigma_{{\rm quark-quark}}$, leaving the rainbow, cloud,
ladder and first exchange graph to $\Sigma_{{\rm quark-gluon}}$. 
This subdivision in  Eqs.(\ref{e:first}) and
(\ref{e:second}) to $J^{(2)}_{{\rm coll}}$ in Eq.(\ref{e:onshell}) will be
handled
separately, as the first term will be seen to lead to elastic quark-quark
and quark-antiquark differential scattering cross sections in the transport
equation, while the $\Sigma_{{\rm quark-gluon}}$ term will be seen to lead
 to processes involving gluons in the
final state, such as the processes 
$q\bar q\rightarrow gg$ and $qg\rightarrow qg$.

\subsubsection{Quark-quark and quark-antiquark scattering cross sections.}

Explicit expressions for the quark loop and its exchange diagram self
energies required in Eq.(\ref{e:first}) are obtained as
\begin{eqnarray}
\Sigma^{(2)+-}_{{\rm q-Loop}} (X,p)&=&-g^4m^4F_{{\rm q-Loop}}^2
\int\!\frac{d^4\!p_1}{(2\pi)^4}
\frac{d^4\!p_2}{(2\pi)^4}\frac{d^4\!p_3}{(2\pi)^4}\frac{d^4\!p_4}{(2\pi)^4}
(2\pi)^4\delta^{(4)}(p-p_1-p_2)\nonumber\\
&&\times (2\pi)^4
\delta^{(4)}(p_2-p_3+p_4)\,S^{+-}(X,p_1)\,G^{++}(X,p_2)\nonumber\\
&&\times S^{+-}(X,p_3)\,S^{-+}(X,p_4)\,G^{--}(X,p_2)
\end{eqnarray}
and
\begin{eqnarray}
\Sigma^{(2)+-}_{E,b)} (X,p)&=&-g^4m^4F_E^2\int\!\frac{d^4\!p_1}{(2\pi)^4}
\frac{d^4\!p_2}{(2\pi)^4}\frac{d^4\!p_3}{(2\pi)^4}\frac{d^4\!p_4}{(2\pi)^4}
(2\pi)^4\delta^{(4)}(p-p_1-p_2)\nonumber\\
&&\times (2\pi)^4
\delta^{(4)}(p_2-p_3+p_4)\,S^{+-}(X,p_1)\,G^{++}(X,p_2)\nonumber\\
&&\times S^{+-}(X,p_3)\,S^{-+}(X,p_4)\,G^{--}(X,p-p_3),
\end{eqnarray}
where $F_{{\rm q-Loop}}$ and $F_E$ are color factors, that will be
given explicitly in the following subsection.   Since they do not
affect our argument, we suppress them in the following.

The integrated collision term, for loss, for example, from
 Eq.(\ref{e:onshell}) can be directly evaluated, to give
the quark loop and exchange contributions
\begin{eqnarray}        
J^{(2)\rm loss}_{\rm coll,q} & = & i g^4m^4\frac{\pi}{E_p}
\int\!\frac{d^4\!p_1}{(2\pi)^4}
\frac{d^4\!p_2}{(2\pi)^4}\frac{d^4\!p_3}{(2\pi)^4}\frac{d^4\!p_4}{(2\pi)^4}
(2\pi)^8\delta^{(4)}(p-p_1-p_2)\delta^{(4)}(p_2-p_3+p_4)\nonumber\\
&&\times\left\{ G^{++}(X,p_2)G^{--}(X,p_2)
         +G^{++}(X,p_2)  G^{--}(X,p-p_3) \right\}\nonumber\\
&&\times(-i\frac{\pi}{E_1})(-i\frac{\pi}{E_3})(-i\frac{\pi}{E_4})
         \sum_{i=1}^8 T_i,
\end{eqnarray}
where
\begin{eqnarray}
T_1&=&\delta(E_1+p_1^0)\,\delta(E_3+p_3^0)\,\delta(E_4+p_4^0)\,f_{\bar{q}}
(-p_1)\,f_{\bar{q}}(-p_3)\,\bar{f}_{\bar{q}}(-p_4)\,f_q(\vec{p})\nonumber\\
T_2&=&\delta(E_1+p_1^0)\,\delta(E_3+p_3^0)\,\delta(E_4-p_4^0)\,f_{\bar{q}}
(-p_1)\,f_{\bar{q}}(-p_3)\,f_q(p_4)\,f_q(\vec{p})\nonumber\\
T_3&=&\delta(E_1+p_1^0)\,\delta(E_3-p_3^0)\,\delta(E_4+p_4^0)\,f_{\bar{q}}
(-p_1)\,\bar{f}_q(p_3)\,\bar{f}_{\bar{q}}(-p_4)\,f_q(\vec{p})\nonumber\\
T_4&=&\delta(E_1+p_1^0)\,\delta(E_3-p_3^0)\,\delta(E_4-p_4^0)\,f_{\bar{q}}
(-p_1)\,\bar{f}_q(p_3)\,f_q(p_4)\,f_q(\vec{p})\nonumber\\
T_5&=&\delta(E_1-p_1^0)\,\delta(E_3+p_3^0)\,\delta(E_4+p_4^0)\,\bar{f}_q(p_1)
\,f_{\bar{q}}(-p_3)\,\bar{f}_{\bar{q}}(-p_4)\,f_q(\vec{p})\nonumber\\
T_6&=&\delta(E_1-p_1^0)\,\delta(E_3+p_3^0)\,\delta(E_4-p_4^0)\,\bar{f}_q(p_1)
\,f_{\bar{q}}(-p_3)\,f_q(p_4)\,f_q(\vec{p})\nonumber\\
T_7&=&\delta(E_1-p_1^0)\,\delta(E_3-p_3^0)\,\delta(E_4+p_4^0)\,\bar{f}_q(p_1)
\,\bar{f}_q(p_3)\,\bar{f}_{\bar{q}}(-p_4)\,f_q(\vec{p})\nonumber\\
T_8&=&\delta(E_1-p_1^0)\,\delta(E_3-p_3^0)\,\delta(E_4-p_4^0)\,\bar{f}_q(p_1)
\,\bar{f}_q(p_3)\,f_q(p_4)\,f_q(\vec{p}).
\end{eqnarray}
One sees that there are eight terms, or eight processes in this 
expression.   However, due
 to energy-momentum-conservation $T_1, T_2, T_4, T_6$ and $T_7$ vanish,
leaving
only $T_3$, $T_5$ and $T_8$.   This is a direct consequence of the on-shell
nature of the quasiparticle approximation.    If this were relaxed, all
terms would necessarily have to be included.    

We can reorganize this expression into a recognizable physical form
by making some simple manipulations.   Letting 
 $p_i \to -p_i$ for the antiquark states and
performing  the $p_1^0$, $p_3^0$, $p_4^0$ and the $p_2$ integration
by absorbing the appropriate $\delta$-functions, we obtain
\begin{eqnarray}
J^{(2)\rm loss}_{\rm coll,q} & = & -g^4m^4\frac{\pi}{E_p}
\int\! \frac{d^3\!p_1}{(2\pi)^32E_1}
\frac{d^3\!p_3}{(2\pi)^3 2E_3}\frac{d^3\!p_4}{(2\pi)^3 2E_4}
(2\pi)^4\nonumber\\
&&\times\left\{\delta^{(4)}(p+p_1-p_3-p_4)\,f_{\bar{q}}
(\vec{p}_1)\,\bar{f}_q(\vec{p}_3)\,\bar{f}_{\bar{q}}(\vec{p}_4)\,
f_q(\vec{p})\right.\nonumber\\
&&\quad\times [ G^{++}(X,p+p_1)G^{--}(X,p+p_1)
               +G^{++}(X,p+p_1)G^{--}(X,p-p_3)]\nonumber\\
&& +\delta^{(4)}(p-p_1+p_3-p_4)\,\bar{f}_q(\vec{p}_1)
   \,f_{\bar{q}}(\vec{p}_3)\,\bar{f}_{\bar{q}}(\vec{p}_4)
   \,f_q(\vec{p})\nonumber\\
&&\quad\times [ G^{++}(X,p-p_1)G^{--}(X,p-p_1)
               +G^{++}(X,p-p_1)G^{--}(X,p+p_3)]\nonumber\\
&& +\delta^{(4)}(p-p_1-p_3+p_4)\,\bar{f}_q(\vec{p}_1)
   \,\bar{f}_q(\vec{p}_3)\,f_q(\vec{p}_4)\,f_q(\vec{p})\nonumber\\
&&\left.\quad\times [ G^{++}(X,p-p_1)G^{--}(X,p-p_1)
               +G^{++}(X,p-p_1)G^{--}(X,p-p_3)] \right\}.
\nonumber \\
\end{eqnarray}
The first two terms of this expression 
can be combined if one makes the substitution $p_1
\leftrightarrow p_3$ in the second term. The third term has a symmetry in
$p_1$ and $p_3$ and can be rewritten as one half the sum of two terms with 
 $p_1$ and $p_3$ interchanged.   The loss term then becomes
\begin{eqnarray}
J^{(2)\rm loss}_{\rm coll,q} & = & -g^4m^4\frac{\pi}{E_p}
\int\! \frac{d^3\!p_1}{(2\pi)^32E_1}
\frac{d^3\!p_3}{(2\pi)^3 2E_3}\frac{d^3\!p_4}{(2\pi)^3 2E_4}
(2\pi)^4\nonumber\\
&&\times\left\{\delta^{(4)}(p+p_1-p_3-p_4)\,f_{\bar{q}}
(\vec{p}_1)\,\bar{f}_q(\vec{p}_3)\,\bar{f}_{\bar{q}}(\vec{p}_4)\,
f_q(\vec{p})\right.\nonumber\\
&&\quad\times [ G^{++}(X,p+p_1)G^{--}(X,p+p_1)
               +G^{++}(X,p+p_1)G^{--}(X,p-p_3)\nonumber\\
&&\quad\quad   +G^{++}(X,p-p_3)G^{--}(X,p-p_3)
               +G^{++}(X,p-p_3)G^{--}(X,p+p_1)]\nonumber\\
&& +\delta^{(4)}(p-p_1-p_3+p_4)\,\bar{f}_q(\vec{p}_1)
   \,\bar{f}_q(\vec{p}_3)\,f_q(\vec{p}_4)\,f_q(\vec{p})\nonumber\\
&&\quad\times \frac{1}{2}[ G^{++}(X,p-p_1)G^{--}(X,p-p_1)
                          +G^{++}(X,p-p_1)G^{--}(X,p-p_3)]\nonumber\\
&&\left.\quad             +G^{++}(X,p-p_3)G^{--}(X,p-p_3)
                          +G^{++}(X,p-p_3)G^{--}(X,p-p_1)]
\right\}.
\nonumber \\
\end{eqnarray}
Using the fact that $[iG^{--}(p)]^{\dagger}=iG^{++}(p)$ and making the 
substitution $p_1
\leftrightarrow p_4$ in the second term, one is able to identify the 
absolute values squared of the Green functions occuring in $J^{(2),{\rm
loss}}_{{\rm coll,q}}$.   One has
\begin{eqnarray}
J^{(2)\rm loss}_{\rm coll,q} & = & g^4m^4\frac{\pi}{E_p}
\int\! \frac{d^3\!p_1}{(2\pi)^32E_1}
\frac{d^3\!p_3}{(2\pi)^3 2E_3}\frac{d^3\!p_4}{(2\pi)^3 2E_4} (2\pi)^4
\delta^{(4)}(p+p_1-p_3-p_4)\nonumber\\
&&\times\left\{\frac{1}{2}\left|iG^{--}(X,p-p_3) +iG^{--}(X,p-p_4)\right|^2
\,f_q(\vec{p}) \,f_q(\vec{p}_1)
\,\bar{f}_q(\vec{p}_3)\,\bar{f}_q(\vec{p}_4)\right.\nonumber\\
&&\quad +\left.\left|iG^{--}(X,p+p_1)+iG^{--}(X,p-p_3)\right|^2
\,f_q(\vec{p}) \,f_{\bar{q}}(\vec{p}_1)
\,\bar{f}_q(\vec{p}_3)\,\bar{f}_{\bar{q}}(\vec{p}_4)\right\}. \nonumber \\
\label{e:jqmid}
\end{eqnarray}
Now one may recognize the scattering amplitude for elastic
quark-quark scattering,
\begin{equation}
-i{\cal M}_{qq\to qq}(p1 \to 34)=(-igm)^2[iG^{--}(p-p_3)+iG^{--}(p-p_4)],
\end{equation}
and the scattering amplitude for quark-antiquark scattering,
\begin{equation}
-i{\cal M}_{q\bar{q}\to q\bar{q}}(p1\to 34)=
(-igm)^2[iG^{--}(p+p_1)+iG^{--}(p-p_3)],
\end{equation}
occurring in Eq.(\ref{e:jqmid}), which may be concisely written as
to give the final result
\begin{eqnarray}
J^{(2)\rm loss}_{\rm coll,q} & = & \frac{\pi}{E_p}
\int\! \frac{d^3\!p_1}{(2\pi)^32E_1}
\frac{d^3\!p_3}{(2\pi)^3 2E_3}\frac{d^3\!p_4}{(2\pi)^3 2E_4} (2\pi)^4
\delta^{(4)}(p+p_1-p_3-p_4)\nonumber\\
&&\times\left\{\frac{1}{2}\left|{\cal M}_{qq\to qq}(p1\to 34)\right|^2
\,f_q(\vec{p}) \,f_q(\vec{p}_1)
\,\bar{f}_q(\vec{p}_3)\,\bar{f}_q(\vec{p}_4)\right.\nonumber\\
&&\quad +\left.\left|{\cal M}_{q\bar{q}\to q\bar{q}}(p1\to 34)\right|^2
\,f_q(\vec{p}) \,f_{\bar{q}}(\vec{p}_1)
\,\bar{f}_q(\vec{p}_3)\,\bar{f}_{\bar{q}}(\vec{p}_4)\right\}.\label{loss1}
\end{eqnarray}
 
The Feynman graphs corresponding to  these
processes are shown in Figs.~8 and 9 respectively.

\subsubsection{Quark-gluon scattering cross sections.}

We now turn our attention to the graphs of 
 Fig.~6.1 to Fig.~6.4(a), which will lead to scattering processes that
involve gluonic degrees of freedom.   As in the previous section,
the Feynman rules for non-equilibrium processes 
 can be applied to these diagrams 
 and the result Wigner transformed. This results in the following
expressions for the self energies, 
\begin{eqnarray}
\Sigma^{(2)+-}_R (X,p)&=&-g^4m^4F_R^2\int\!\frac{d^4\!p_1}{(2\pi)^4}
\frac{d^4\!p_2}{(2\pi)^4}\frac{d^4\!p_3}{(2\pi)^4}\frac{d^4\!p_4}{(2\pi)^4}
(2\pi)^4\delta^{(4)}(p-p_1-p_2)\nonumber\\
&&\times (2\pi)^4
\delta^{(4)}(p_2-p_3-p_4)\,G^{+-}(X,p_1)\,S^{++}(X,p-p_3)\nonumber\\
&&\times G^{+-}(X,p_3)\,S^{+-}(X,p_4)\,S^{--}(X,p-p_3)
\end{eqnarray}
for the rainbow diagram,
\begin{eqnarray}
\Sigma^{(2)+-}_L (X,p)&=&-\frac{1}{2}g^4m^4F_L^2\int\!\frac{d^4\!p_1}{(2\pi)^4}
\frac{d^4\!p_2}{(2\pi)^4}\frac{d^4\!p_3}{(2\pi)^4}\frac{d^4\!p_4}{(2\pi)^4}
(2\pi)^4\delta^{(4)}(p-p_1-p_2)\nonumber\\
&&\times (2\pi)^4
\delta^{(4)}(p_2-p_3-p_4)\,G^{+-}(X,p_1)\,G^{++}(X,p_2)\nonumber\\
&&\times G^{+-}(X,p_3)\,S^{+-}(X,p_4)\,G^{--}(X,p_2) 
\label{ladder}
\end{eqnarray}
for the ladder graph,
\begin{eqnarray}
\Sigma^{(2)+-}_{C,(a)/(b)} (X,p)&=&-g^4m^4F_C^2\int\!\frac{d^4\!p_1}{(2\pi)^4}
\frac{d^4\!p_2}{(2\pi)^4}\frac{d^4\!p_3}{(2\pi)^4}\frac{d^4\!p_4}{(2\pi)^4}
(2\pi)^4\delta^{(4)}(p-p_1-p_2)\nonumber\\
&&\times (2\pi)^4
\delta^{(4)}(p_2-p_3-p_4)\,G^{+-}(X,p_1)\,G^{\pm\pm}(X,p_2)\nonumber\\
&&\times G^{+-}(X,p_3)\,S^{+-}(X,p_4)\,S^{\mp\mp}(X,p-p_3) 
\end{eqnarray}
for the two cloud diagrams, and
\begin{eqnarray}
\Sigma^{(2)+-}_{E,(a)} (X,p)&=&-g^4m^4F_E^2\int\!\frac{d^4\!p_1}{(2\pi)^4}
\frac{d^4\!p_2}{(2\pi)^4}\frac{d^4\!p_3}{(2\pi)^4}\frac{d^4\!p_4}{(2\pi)^4}
(2\pi)^4\delta^{(4)}(p-p_1-p_2)\nonumber\\
&&\times (2\pi)^4
\delta^{(4)}(p_2-p_3-p_4)\,G^{+-}(X,p_1)\,S^{++}(X,p-p_4)\nonumber\\
&&\times G^{+-}(X,p_3)\,S^{+-}(X,p_4)\,S^{--}(X,p-p_3) 
\end{eqnarray}
for the first exchange diagram.  $F_R$, $F_L$, $F_C$ and $F_E$ are 
appropriate color factors, that will be discussed in detail in the following
subsection, but which will be suppressed here.
Note that a  factor $1/2$ occurs  in the expression for the ladder diagram 
because of the gluon loop.
The expressions for $\Sigma^{(2)-+}$ are obtained from the
ones for $\Sigma^{(2)+-}$ by exchanging $-$ and $+$.
To construct the integrated 
 collision term, one has as in the previous section
 to integrate over $p^0$ over
an
interval $\Delta_{\pm}$ which contains $\pm E_p(X)$ and one finds, using
Eq.(\ref{e:onshell}) that the loss term of the collision
incorporating the rainbow, cloud, ladder and first exchange graphs,
is given as
\begin{eqnarray}        
J^{(2)\rm loss}_{\rm coll,g} & = & i g^4m^4\frac{\pi}{E_p}
\int\!\frac{d^4\!p_1}{(2\pi)^4}
\frac{d^4\!p_2}{(2\pi)^4}\frac{d^4\!p_3}{(2\pi)^4}\frac{d^4\!p_4}{(2\pi)^4}
(2\pi)^8\delta^{(4)}(p-p_1-p_2)\delta^{(4)}(p_2-p_3-p_4)\nonumber\\
&&\times\left\{ S^{++}(X,p-p_3)S^{--}(X,p-p_3)
               +\frac{1}{2}G^{++}(X,p_2)  G^{--}(X,p_2)\right.\nonumber\\
&&\quad        + G^{++}(X,p_2)S^{--}(X,p-p_3)
               + G^{--}(X,p_2)S^{++}(X,p-p_3)\nonumber\\
&&\left.\quad  + S^{++}(X,p-p_4)S^{--}(X,p-p_3) \right\}
        (-i\frac{\pi}{E_1})(-i\frac{\pi}{E_3})(-i\frac{\pi}{E_4})
         \sum_{i=1}^8 T_i,
\end{eqnarray}        
  where       
\begin{eqnarray}        
T_1&=&\delta(E_1+p_1^0)\,\delta(E_3+p_3^0)\,\delta(E_4+p_4^0)\,f_{\bar{q}}
(-p_1)\,f_{\bar{g}}(-p_3)\,f_{\bar{g}}(-p_4)\,f_q(\vec{p})\nonumber\\
T_2&=&\delta(E_1+p_1^0)\,\delta(E_3+p_3^0)\,\delta(E_4-p_4^0)\,f_{\bar{q}}
(-p_1)\,f_{\bar{g}}(-p_3)\,\bar{f}_g(p_4)\,f_q(\vec{p})\nonumber\\
T_3&=&\delta(E_1+p_1^0)\,\delta(E_3-p_3^0)\,\delta(E_4+p_4^0)\,f_{\bar{q}}
(-p_1)\,\bar{f}_g(p_3)\,f_{\bar{g}}(-p_4)\,f_q(\vec{p})\nonumber\\
T_4&=&\delta(E_1+p_1^0)\,\delta(E_3-p_3^0)\,\delta(E_4-p_4^0)\,f_{\bar{q}}
(-p_1)\,\bar{f}_g(p_3)\,\bar{f}_g(p_4)\,f_q(\vec{p})\nonumber\\
T_5&=&\delta(E_1-p_1^0)\,\delta(E_3+p_3^0)\,\delta(E_4+p_4^0)\,\bar{f}_q(p_1)
\,f_{\bar{g}}(-p_3)\,f_{\bar{g}}(-p_4)\,f_q(\vec{p})\nonumber\\
T_6&=&\delta(E_1-p_1^0)\,\delta(E_3+p_3^0)\,\delta(E_4-p_4^0)\,\bar{f}_q(p_1)
\,f_{\bar{g}}(-p_3)\,\bar{f}_g(p_4)\,f_q(\vec{p})\nonumber\\
T_7&=&\delta(E_1-p_1^0)\,\delta(E_3-p_3^0)\,\delta(E_4+p_4^0)\,\bar{f}_q(p_1)
\,\bar{f}_g(p_3)\,f_{\bar{g}}(-p_4)\,f_q(\vec{p})\nonumber\\
T_8&=&\delta(E_1-p_1^0)\,\delta(E_3-p_3^0)\,\delta(E_4-p_4^0)\,\bar{f}_q(p_1)
\,\bar{f}_g(p_3)\,\bar{f}_g(p_4)\,f_q(\vec{p}).
\end{eqnarray}        
Once again, eight terms result from this multiplication.   Now, again
due to energy-momentum-conservation, $T_1, T_2, T_3, T_5$ and $T_8$ vanish,
and we are left with three non-vanishing terms, $T_4$, $T_6$ and $T_7$.
 
Applying the same procedure as for $J^{(2)\rm loss}_{\rm coll,q}$
as in the previous subsection, one can regroup the remaining terms to
read
\begin{eqnarray}
J^{(2)\rm loss}_{\rm coll,g} & = & g^4m^4\frac{\pi}{E_p}
\int\! \frac{d^3\!p_1}{(2\pi)^32E_1}
\frac{d^3\!p_3}{(2\pi)^3 2E_3}\frac{d^3\!p_4}{(2\pi)^3 2E_4} (2\pi)^4
\delta^{(4)}(p+p_1-p_3-p_4)\nonumber\\
&&\times\left\{\frac{1}{2}\left|iG^{--}(X,p+p_1) + iS^{--}(X,p-p_3)
+ iS^{--}(X,p-p_4)\right|^2\right.\nonumber\\
&&\quad\times f_q(\vec{p}) \,f_{\bar{q}}(\vec{p}_1)
\,\bar{f}_g(\vec{p}_3)\,\bar{f}_g(\vec{p}_4)\nonumber\\
&&\quad +\left|iS^{--}(X,p+p_1) + iG^{--}(X,p-p_3) +
iS^{--}(X,p-p_4)\right|^2\nonumber\\
&&\quad \left.\times f_q(\vec{p}) \,f_g(\vec{p}_1)
\,\bar{f}_q(\vec{p}_3)\,\bar{f}_g(\vec{p}_4)\right\}.
\label{e:jgmid}
\end{eqnarray}
In order to identify the physical processes that give rise to these terms,
we examine first all possible contributions to the annihilation process 
$q\bar q\rightarrow gg$.   The Feynman graphs for this within this model
are shown in Fig.~10.   
The scattering amplitude associated therewith is 
\begin{equation}
-i{\cal M}_{q\bar{q}\to gg}(p1\to 34)=(-igm)^2[iG^{--}(p+p_1) +
iS^{--}(p-p_3) + iS^{--}(p-p_4)].
\label{e:stu1}
\end{equation}
In a similar manner, the elastic scattering process
 $qg\to qg$, which is shown in Fig.~11, has the scattering
amplitude 
\begin{equation}
-i{\cal M}_{qg\to qg}(p1\to 34)=(-igm)^2[iS^{--}(p+p_1) +
iG^{--}(p-p_3) + iS^{--}(p-p_4)].
\label{e:stu2}
\end{equation}
 
One can identify the absolute value squared of Eqs.(\ref{e:stu1}) and
(\ref{e:stu2}) in Eq.(\ref{e:jgmid}) and
therefore $J^{(2)\rm loss}_{\rm coll,g}$ can be written as
\begin{eqnarray}
J^{(2)\rm loss}_{\rm coll,g} & = & \frac{\pi}{E_p}
\int\! \frac{d^3\!p_1}{(2\pi)^32E_1}
\frac{d^3\!p_3}{(2\pi)^3 2E_3}\frac{d^3\!p_4}{(2\pi)^3 2E_4} (2\pi)^4
\delta^{(4)}(p+p_1-p_3-p_4)\nonumber\\
&&\times\left\{\frac{1}{2}\left|{\cal M}_{q\bar{q}\to gg}(p1\to 34)\right|^2
\,f_q(\vec{p}) \,f_{\bar{q}}(\vec{p}_1)
\,\bar{f}_g(\vec{p}_3)\,\bar{f}_g(\vec{p}_4)\right.\nonumber\\
&&\quad +\left.\left|{\cal M}_{qg\to qg}(p1\to 34)\right|^2
\,f_q(\vec{p}) \,f_g(\vec{p}_1)
\,\bar{f}_q(\vec{p}_3)\,\bar{f}_g(\vec{p}_4)\right\}.\label{loss2}
\end{eqnarray}
The complete loss term is obtained by adding Eq.(\ref{loss1}) and
(\ref{loss2}),
\begin{equation}
J^{(2) {{\rm loss}}}_{{\rm coll}} = J^{(2) {{\rm loss}}}_{{\rm coll,q
}} + J^{(2) {{\rm loss}}}_{{\rm coll,g}}.
\label{e:end}
\end{equation}
The gain term can be constructed by replacing $f
\leftrightarrow \bar{f}$ in the complete loss term. 
With the relation 
\begin{equation}
\frac{d\sigma}{d\Omega} = \frac{|{\cal M}|^2}{|\vec v_p-\vec v_1|2E_p 2E_1}
\frac{dQ}{d\Omega}
\end{equation}
and the phase space factor
\begin{equation}
Q = (2\pi)^4 \delta^{(4)}(p+p_1-p_3-p_4) \frac{d^3p_3}{(2\pi)^32E_3}
\frac{d^3p_4}{(2\pi)^32E_4},
\end{equation}
the final form for the Boltzmann equation, calculated to second order in the 
number of exchanged gluons, is for quarks ($a=q$)
\begin{eqnarray}
&&2p \partial_X f_a(X,\vec p) 
 =\int\! d\Omega\frac{d^3p_1}{(2\pi)^3 2E_1} 
|\vec v_p-\vec v_1| 2E_p 2E_1 \nonumber \\
&&\times\sum_{j=1}^4 s_j\frac{d\sigma_j}{d\Omega}\bigg|_{ab \to cd}
\left[\bar f_a(\vec p)\bar f_b(\vec p_1) f_c(\vec p_3) 
 f_d(\vec p_4)  -   f_a(\vec p) f_b(\vec p_1)
\bar f_c(\vec p_3)\bar f_d(\vec p_4)\right],
\label{e:bfinal}
\end{eqnarray}
where partons $b$, $c$, and $d$ can be a quark, antiquark or gluon,
and $j$ labels the four processes $j=1...4$ corresponding to $q\bar q
\rightarrow gg$, $qg\rightarrow qg$ , $qq\rightarrow qq$ and $q\bar
q\rightarrow q\bar q$.   The $s_j$ are symmetry factors $s_1=s_3=1/2$ and
$s_2=s_4=1$.
 
The transport equation for gluons can be obtained in
an analogous way and calculated for two loops, it takes  
the same form as Eq.(\ref{e:bfinal}) with $a=g$.
Then $j$ labels the four processes $j=1...4$ corresponding to $gg\to gg$,
$gg\to q\bar q$, $gq\to gq$ and $g\bar q\to g\bar q$. The appropriate
symmetry factors are $s_1=1/2$ and $s_2=s_3=s_4=1$.

\subsection{Color expansion}

 We now deal with the color factors, which we have neglected so far.
We calculate them for \emph{one} $SU(N)$ color group. The overall color
factor for both color groups is then obtained by squaring it.
 
The matrices $(t^a)_{ij}$ are the matrices of the color group in the
representation of the quarks, while $(T^a)_{bc}=-if_{abc}$ are the color
matrices in the adjoint representation and $f_{abc}$ are the structure
constants of the color group, see Eq.(\ref{e:commutator}).
The $t^a$'s are normalized to
\begin{equation}
{\rm tr}(t^at^b) = \frac{1}{2} \delta_{ab}.
\end{equation}                        
The ``square'' of the generator in some representation must be proportional to
the unit operator (Schur's Lemma). Therefore
\begin{equation}                       
(t^a)_{ij}(t^a)_{jk} = C_F \delta_{ik}
\end{equation}                        
and                        
\begin{equation}                       
T^a_{bd}T^a_{dc} = f_{bad}f_{cad} = C_A \delta_{bc},
\end{equation}                        
where the numbers $C_F$ and $C_A$ are the Casimir operators of the
fundamental and adjoint representation, respectively.
They take the values (see for example \cite{dokshi})
\begin{equation}                       
C_F = \frac{N_c^2-1}{2N_c} 
\end{equation}                        
and                        
\begin{equation}                       
C_A = N_c.                 
\end{equation}                        
Consider now the quark self energies that were evaluated in the previous
subsection.   Let $i$ denote the external parton index.   It is therefore not
to be summed over.
The color factor for the rainbow graph is
\begin{equation}                       
F_R = t^b_{ij} t^a_{jk} t^a_{kl} t^b_{li} = C_F^2 \delta_{ii}
= \frac{(N_c^2-1)^2}{4N_c^2}\delta_{ii}.
\end{equation}                        
For the ladder graph, one finds
\begin{equation}                       
F_L = (-if_{abc})(-if_{cbd})t^a_{ij} t^d_{ji}
= C_A \delta_{ad}t^a_{ij} t^d_{ji} = C_A C_F \delta_{ii}
= \frac{N_c^2-1}{2}\delta_{ii}.
\end{equation}                        
For the cloud graph, one obtains
\begin{equation}                       
F_C = (-if_{acb})t^a_{ij} t^b_{jk} t^c_{ki} = -\frac{N_c^2-1}{4}\delta_{ii},
\end{equation}                        
where the relation (see for example \cite{dokshi})
\begin{equation}                       
-if_{abc}t^a t^b = \frac{C_A}{2}t^c
\end{equation}                        
has been  used.                  
The color factor for the exchange graph is
\begin{equation}                       
F_E = t^a_{ij} t^b_{jk} t^a_{kl} t^b_{li} =
-\frac{N_c^2-1}{4N_c^2}\delta_{ii},
\end{equation}                        
where the relation \cite{dokshi}
\begin{equation}                       
t^a t^b t^a = \frac{-1}{2N_c} t^b
\end{equation}                        
has been used.                  
Finally, for the quark loop graph the color factor is given by
\begin{equation}                       
F_{{\rm q-Loop}} = t^a_{ij} t^b_{ji}{\rm tr}(t^a t^b) =
\frac{N_c^2-1}{4N_c}\delta_{ii}.
\end{equation}     
Up to this point, we have made a semiclassical expansion that involves
keeping only the leading term in expanding the exponential in
Eq.(\ref{lambda1})  (here the factor $\hbar$ has been set to one.)
In addition, we have examined sets of diagrams organized according to the 
number of interaction lines, i.e. according to the coupling strength.   We
have
found that all graphs in a class are required in order to build up the cross
sections that ultimately occur in a Boltzmann like equation.    However,
at the level of two exchanged gluons, we are already faced with five types
of graphs, and this number increases rapidly with the number of exchanged
gluons.   One possible simplifying assumption is the additional imposition
of an expansion in the inverse number of colors.   According to such a
criterion, the ladder, the rainbow and the cloud diagrams are leading, 
since their color factors for one color group 
are of order $O(N_c^2)$ while for the quark-loop diagram it goes as  
$N_c$ and for the  exchange diagram only as $N_c^0$.

Since the ladder and the rainbow diagram lead to cross sections involving
gluons while the quark-loop diagram leads to elastic 
quark-(anti)quark cross sections, one can conclude that the quark degrees of
freedom are suppressed in comparison with the gluon degrees of freedom.
This is in agreement with the results of an evaluation of the 
quark-quark scattering amplitude within this
model \cite{forsh}, in which the quark degrees of freedom are neglected,
however due
to kinematical reasons.

Although the ladder and the rainbow diagram are both of order $O(N_c^2)$,
the ratio of their color factors for one color group is not one, but
\begin{equation}
\frac{F_R}{F_L}=\frac{C_F}{C_A},
\end{equation}
which is $4/9 \approx 1/2$ for $N_c=3$.
Since in the rainbow diagram the second gluon couples at the quark line
while in the ladder diagram it couples at the first gluon,
two quark-quark-gluon vertices are suppressed by a factor $4/9$ per color
group in
comparison with two 3-gluon vertices.   Thus,  there is no {\it strict}
ordering of the gluon graphs according to a single class of diagrams, in an
expansion in $1/N_c$.
Although  the ladder graphs and the processes that 
they lead to appear largest, one should not that the symmetry factors of the
other graphs compensate for this.   A numerical study is essential to
determine the actual order of magnitude of each graph.

As a final comment, note that an expansion in color also incorporates the
coupling strength.   Assuming that $g\sim 1/N_c$, we find that the Fock
term $\sim g^2N_c^2$ and the ladder diagram $\sim g^4N_c^4$ are of the
same order.

\subsection{Three parton production}

Since many diagrams contribute to the self energy with three exchanged
gluons, we again make an additional expansion in the inverse number of 
colors as explained in the previous Subsection, and consider only diagrams 
that do not contain quark loops.
In the case of two exchanged gluons, we have seen that 
 the ladder and the rainbow diagrams give
rise to the squared scattering amplitudes of \emph{individual} channels. 
Therefore, we conclude (and will check later) that for the case with three 
exchanged gluons, the three-rung ladder diagram, the rainbow diagram with 
three bows and all possible mixtures between them will lead to the squared
scattering amplitudes of \emph{individual} channels of processes involving
three gluons. The generic form of these diagrams is given in 
Fig.~12. 
(Note: diagrams that would generate crossed terms
 between the single channels are shown in Appendix C).

A direct evaluation of the self energy diagrams in Fig.~12  should give rise to
cross sections of processes that involve three gluons, {\it i.e.}
one expects, for example, the three gluon production $q\bar{q} \to ggg$ but
also multiparton production such as  
$qg\to qgg$. The generic Feynman diagrams for these processes are shown in 
Figs.~13 and 14 respectively.

To be specific, let us consider now the ladder diagram
 of Fig.~12(a), in particular 
 $\Sigma^{(3)+-}_L(X,p)$. There are sixteen possibilities of arranging 
the signs that can be 
associated with the vertices of the first and second rungs.
Invoking the on-shell argument and energy-momentum conservation, 
the possibilities become restricted 
to the single diagram shown in Fig.~15.
The explicit expression for this ladder diagram is
\begin{eqnarray}
\Sigma^{(3)+-}_L (X,p)&=&-i\frac{1}{2}g^6m^6\int\!\frac{d^4\!p_1}{(2\pi)^4}
\frac{d^4\!p_2}{(2\pi)^4}\frac{d^4\!p_3}{(2\pi)^4}\frac{d^4\!p_4}{(2\pi)^4}
\frac{d^4\!p_5}{(2\pi)^4}\frac{d^4\!p_6}{(2\pi)^4}\nonumber\\
&&\times(2\pi)^4\delta^{(4)}(p-p_1-p_2)(2\pi)^4\delta^{(4)}(p_2-p_3-p_4)
(2\pi)^4\delta^{(4)}(p_4-p_5-p_6)\nonumber\\
&&\times S^{+-}(X,p_1)\,G^{++}(X,p_2)\,G^{+-}(X,p_3)
\,G^{++}(X,p_4)\nonumber\\
&&\times G^{+-}(X,p_5)\,G^{+-}(X,p_6)\,G^{--}(X,p_4)\,G^{--}(X,p_2).
\label{e:ladder3}
\end{eqnarray}

Once again, one may insert the quasiparticle {\it ans\"atze} for the quark
and gluonic Green functions from Eqs.(\ref{d-+}) - (\ref{d++}), and
construct the contribution of this ladder term to $J^{(3){\rm loss}}_{{\rm
coll}}$ as in Subsection IV A by taking its on-shell value
\begin{equation}
J^{(3){\rm loss}}_{{\rm coll}} 
= -i\frac{\pi}{E_p}\Sigma^{(3)+-}(X,p_0=E_p,\vec p) f_q(X,\vec p).
\label{e:j3loss}
\end{equation}
Multiplying out the product of Green functions in terms of the quark and
gluonic distribution functions enables us to identify all processes that
can occur.   The details of this cumbersome calculation are given in
Appendix B.   Here we simply note that the constraints of energy-momentum
conservation 
admit all processes which have at least two partons in both
the initial and final state.   That is to say, in addition to the parton
production processes $q\bar q\to ggg$ and $qg\to qgg$, that were already
mentioned, one also obtains the matrix elements for the processes $
q\bar qg\to gg$ and $qgg\to qg$.

In order to complete the calculation, the remaining diagrams of Fig.~12
and additional self energy graphs that lead to mixed terms of the scattering
amplitudes must be calculated in an analogous way.   There are very many
such 
diagrams that build the mixed terms between the scattering amplitudes 
that were shown in Figs.~13 and 14.     For completeness, we show these
in Appendix C.   We will not however calculate these graphs
in any detail here.

The final form of the Boltzmann equation, calculated to third order in the
number of exchanged gluons, reads
\begin{eqnarray}
&&2p \partial_X f_q(X,\vec p)
=\int\! \frac{d^3p_1}{(2\pi)^3 2E_1}\frac{d^3p_2}{(2\pi)^3
2E_2}\frac{d^3p_3}{(2\pi)^3 2E_3}(2\pi)^4\nonumber\\
&&\times\left\{ \delta^{(4)}(p+p_1-p_2-p_3)\sum_{j=1}^4 s_j
\left|{\cal M}_j(qa \to bc)\right|^2\right.\nonumber\\
&&\quad\times\left[\bar f_q(\vec p)\bar f_a(\vec p_1) f_b(\vec p_3)
 f_c(\vec p_4)  -   f_q(\vec p) f_a(\vec p_1)
\bar f_b(\vec p_3)\bar f_c(\vec p_4)\right]  \nonumber\\
&&\quad +\int\! \frac{d^3p_4}{(2\pi)^3 2E_4}
\left\{   \delta^{(4)}(p+p_1-p_2-p_3-p_4) \sum_{j=5}^6 s_j 
\left|{\cal M}_j(qa \to bcd)\right|^2   \right.  \nonumber\\
&&\qquad\qquad\qquad\quad \times \left[ \bar f_q(\vec p)\bar f_a(\vec p_1)
f_b(\vec p_3)f_c(\vec p_4)f_d(\vec p_5) - f_q(\vec p) f_a(\vec p_1)
\bar f_b(\vec p_3)\bar f_c(\vec p_4)\bar f_d(\vec p_5)\right]
\nonumber \\
&& \qquad \qquad\qquad\quad+ \delta^{(4)}(p+p_1+p_2-p_3-p_4) \sum_{j=7}^8 s_j
\left|{\cal M}_j(qab \to cd)\right|^2 \nonumber\\
&&\qquad\qquad\qquad\quad \times \left.\left[ \bar f_q(\vec p)\bar f_a(\vec p_1)
\bar f_b(\vec p_3)f_c(\vec p_4)f_d(\vec p_5) - f_q(\vec p) f_a(\vec p_1)
f_b(\vec p_3)\bar f_c(\vec p_4)\bar f_d(\vec p_5)\right]\right\},
\nonumber \\
\label{e:bfinal3}
\end{eqnarray}
where $a$, $b$, $c$ and $d$ are partons and 
$j=1...4$ labels the four processes given after Eq.(\ref{e:bfinal}) while 
$j=5...8$ labels the four processes corresponding to $q\bar q \to
ggg$, $qg\to qgg$, $q\bar qg\to gg$ and $qgg\to qg$. The symmetry factors
for $j=5...8$ are given as $s_5=1/3!$ and $s_6=s_7=s_8=1/2$.

\subsection{$n$ parton production}

Of the self energy diagrams of order $O(g^{2 \rm n})$, the ladder diagram, 
the rainbow diagram and all possible mixtures between these two are leading 
in an expansion in $1/N_c$. On evaluation, 
these diagrams lead to the scattering process $q\bar q\to {\rm n}g$ and 
all possible crossed processes, such as $q\bar qg\to ({n}-1)g$, $qg\to q 
({n}-1)g$, ... in which  at least two partons occur both in the initial and  
final states. 
The Boltzmann equation for quarks then reads 
\begin{eqnarray}
&&p \partial_X f_q(X,\vec p)
 =\frac \pi{E_p}\sum_{m,n=1}^{\infty}
\int\! \frac{d^3k}{(2\pi)^3 2E_k}\frac{d^3p_1}{(2\pi)^3 2E_1}
...\frac{d^3p_{m+n}}{(2\pi)^3 2E_{m+n}}(2\pi)^4\nonumber\\
&&\times \left\{ 
\delta^{(4)}(p+k+p_1+...+p_{m-1}-p_m-...-p_{m+n})\right.\nonumber\\
&&\quad \times s_{m-1} s_{n+1}\,
\left|{\cal M}(q\bar q \,(m-1)g \to (n+1)g)\right|^2\nonumber\\
&&\quad\times \left[\bar f_q(\vec p)\bar f_{\bar q}(\vec k)\bar f_g(\vec p_1)
...\bar f_g(\vec p_{m-1}) f_g(\vec{p}_m)... f_g(\vec{p}_{m+n})\right.\nonumber\\
&&\left.\quad\; - f_q(\vec p) f_{\bar q}(\vec k)f_g(\vec p_1)...f_g(\vec
p_{m-1})\bar f_g(\vec p_m) ...\bar f_g(\vec p_{m+n})\right] \nonumber\\
&& + \;\delta^{(4)}(p+p_1+...+p_m-k-p_{m+1}-...-p_{m+n})
s_{m}s_{n}\,\left|{\cal M}(q\,mg \to q\,ng)\right|^2
\nonumber\\
&&\quad\times \left[\bar f_q(\vec p)\bar f_g(\vec p_1)...
  \bar f_g(\vec p_{m})f_q(\vec k) f_g(\vec{p}_{m+1})...f_g(\vec{p}_{m+n})
\right.\nonumber\\
&&\quad\;\left.   - f_q(\vec p) f_g(\vec p_1)...f_g(\vec p_{m})
   \bar f_q(\vec k) \bar f_g(\vec{p}_{m+1}) ...\bar f_g(\vec{p}_{m+n})
\right] \Bigg\} \nonumber\\
\label{e:bfinaln}
\end{eqnarray}
with the symmetry factors $s_n=1/n!$, which is our final result.

\section{Summary and conclusions}
\label{sec:summary}

In this paper, we have worked out the transport theory for the scalar
partonic model.   We have placed particular emphasis on the collision term
examining the collision integral first in the mean field (Hartree and Fock
approximations) and then including more loops.
On making an analysis of the color structure, we have found that a $1/N_c$
expansion leads to a suppression of the quark loops and their exchange
diagrams, much as occurs in QCD.   Thus it turns out that all remaining
self energy diagrams that lead to gluon production and quark-gluon
scattering are of the same order.    We have not been able to find any 
convincing argument to favor an expansion that only includes ladder graphs,
as is for example the dominant contribution in elastic quark-quark
scattering in the Regge limit, when considered on its own within this model.

  This can
simply be traced back to the fact that the uppermost line in the quark or
gluon  self
energy
is closed and therefore possesses a momentum that must be integrated over.
This fact reemerges in the Boltzmann equation in that all momenta except
for that of the particle under study, are integrated over.  An additional 
assumption, e.g. that the uppermost gluonic momentum should take on a 
specific value, is required in order to generate a leading logarithmic
expansion.

We have also analysed the situation involving three loops, which leads
to three parton production, and find that initial state interactions also
enter into the picture.  Finally, a generalized Boltzmann like equation
that includes multiple gluon production, but which also incorporates 
initial state interactions, is given.

In concluding, we comment that further study is required in order to
establish the factors that could lead to mass renormalization in this model
and which has not yet been addressed.

\section{Acknowledgments}
\label{sec:ack}
We wish to thank Klaus Werner at the Ecole des Mines de Nantes for many
fruitful discussions and his kind hospitality at this institute while
this project was discussed.

\begin{appendix}
\section{Green functions, transport and constraint equations}
In this appendix, we give a brief guideline for the derivation of the
transport and constraint equations, Eqs.(\ref{transport}) and
(\ref{constraint}).

The Schwinger-Keldysh Green functions defined in Eqs.(\ref{e:defs}) and
(\ref{e:defg})
can be summarized in a compact 
matrix form 
\begin{equation}
\underline{D}={D^{--}\;\,D^{-+}\choose D^{+-}\;\,D^{++}},
\end{equation}
using the generic notation already introduced in Section IIB.
From the definition of the Green functions,  the relation
\begin{equation}
D^{--}(x,y)+D^{++}(x,y)=D^{-+}(x,y)+ D^{+-}(x,y)
\end{equation}
follows,
showing that the four components $D^{ij}$ are not independent.
We define the retarded and advanced Green functions in the standard way as 
\begin{eqnarray}
D^R(x,y)&:=&\;\;\,\Theta(x_0-y_0)[D^{+-}(x,y)-D^{-+}(x,y)]\nonumber\\ 
        &=& D^{--}(x,y) - D^{-+}(x,y) = D^{+-}(x,y) - D^{++}(x,y)\label{sr}\\ 
D^A(x,y)&:=&-\Theta(y_0-x_0)[D^{+-}(x,y)-D^{-+}(x,y)]\nonumber\\
        &=& D^{--}(x,y) - D^{+-}(x,y) = D^{-+}(x,y) - D^{++}(x,y).\label{sa}
\end{eqnarray}
The equations of motion that the Green functions satisfy are
\begin{equation}
(\Box_x + M^2)\underline{D}(x,y)=-\underline{\sigma}_z\delta^{(4)}(x-y)+
\int d^4z\,\underline{\sigma}_z\,\underline{\Pi}(x,z)\underline{D}(z,y),
\label{eomgreen}
\end{equation}
given in terms of  the irreducible proper self energy 
\begin{equation}
\underline{\Pi}={\Pi^{--}\,\;\Pi^{-+}\choose 
\Pi^{+-}\,\;\Pi^{++}}
\end{equation}
and  
\begin{equation}
\underline{\sigma}_z={1\;\;\;\;0\choose 0\,-\!1}.
\end{equation}
In Eq.(\ref{eomgreen}), $M$ is the free bosonic parton mass.

The four components of the self energy are also not independent. From their
definition,  the relation 
\begin{equation}
\Pi^{--} + \Pi^{++} = -( \Pi^{+-} + \Pi^{-+})
\end{equation}
can be seen to hold.
The retarded and  advanced self energies are defined to be
\begin{eqnarray}
\Pi^R(x,y) &=& \Pi^{--}(x,y) + \Pi^{-+}(x,y)\label{sigr}\nonumber\\
\Pi^A(x,y) &=& \Pi^{--}(x,y) + \Pi^{+-}(x,y).\label{siga}
\end{eqnarray}
We now consider specifically the equation of motion  for $D^{-+}$.   This reads
\begin{eqnarray}
(\Box_x+ M^2) D^{-+}(x,y)&=&\int d^4z\{\Pi^{--}(x,z)D^{-+}(z,y)+
\Pi^{-+}(x,z)D^{++}(z,y)\}\nonumber\\
&=&\int d^4 z \{ \Pi^A(x,z)D^{-+}(z,y)-\Pi^{+-}(x,z)D^{-+}(z,y)\nonumber\\
&&\quad\quad+\Pi^{-+}(x,z)D^{+-}(z,y)-\Pi^{-+}(x,z)D^R(z,y)\}
\end{eqnarray}
while the  conjugate equation is
\begin{eqnarray}     
({\Box}_y+M^2) D^{-+}(x,y)&=&-\int d^4z\{D^{-+}(x,z)\Pi^{++}(z,y)
+D^{--}(x,z)\Pi^{-+}(z,y)\}\nonumber\\                       
&=&\int d^4 z \{ D^{-+}(x,z)\Pi^A(z,y)-D^R(x,z)\Pi^{-+}(z,y)\}.
\end{eqnarray}    
Now a Wigner transform of both equations is performed to yield
\begin{eqnarray} 
&&\left[ \frac{1}{4}\Box_X -ip\partial_X-p^2+M^2\right]D^{-+}(X,p)=\nonumber
\\
&&\quad\quad
\;\;\,\Pi^A(X,p)\hat{\Lambda}D^{-+}(X,p)
-\Pi^{+-}(X,p)\hat{\Lambda}D^{-+}(X,p)\nonumber\\
&&\quad\quad +\Pi^{-+}(X,p)\hat{\Lambda}D^{+-}(X,p)
-\Pi^{-+}(X,p)\hat{\Lambda}D^R(X,p)
\label{ad1} 
\end{eqnarray}
and
\begin{eqnarray}
&&\left[ \frac{1}{4}{\Box}_X 
+ip{\partial}_X - p^2+M^2\right]D^{-+}(X,p)=\nonumber \\
&&\quad\quad 
 S^{-+}(X,p)\hat{\Lambda}\Sigma^A(X,p)-S^R(X,p)\hat{\Lambda}\Sigma^{-+}(X,p),
\nonumber \\
\label{ad2}
\end{eqnarray}
with the differential operator 
\begin{equation}
\hat{\Lambda}:={\rm exp}\left\{ \frac{-i}{2}\left(\overleftarrow{\partial}_X
 \overrightarrow{\partial}_p
-\overleftarrow{\partial}_p\overrightarrow{\partial}_X \right)\right\}.
\label{lambda}
\end{equation}
Subtracting Eq.(\ref{ad1}) from Eq.(\ref{ad2}) 
gives the so-called transport equation,
\begin{equation}
-2ip\partial_X D^{-+}(X,p)=I_-,
\end{equation}
while adding them yields the so-called constraint equation,
\begin{equation}
\left( \frac{1}{2} \Box_X - 2p^2 +2M^2\right) D^{-+}(X,p) = I_+,
\end{equation}
that were quoted as Eqs.(\ref{transport}) and (\ref{constraint}), and
$I_\mp$ is as given in Eqs.(\ref{e:imp}) to (\ref{e:pim}).

\section{Three parton production}

In this Appendix, the explicit evaluation of the ladder graph for three
parton production is given. 
 On inserting the quasiparticle Green functions from Eqs.(\ref{d-+}) to
(\ref{d++}) in Eq.(\ref{e:ladder3}) and 
after performing the $p_2$, $p_4$, $p_1^0$, $p_3^0$, $p_5^0$ and
$p_6^0$ integration by absorbing the appropriate $\delta$-functions
we obtain from Eq.(\ref{e:j3loss}) the result
\begin{eqnarray}
J^{(3){\rm loss}}_{{\rm coll}} &=& -\frac{1}{2}g^6m^6\frac{\pi}{E_p}
\int \! \frac{d^3p_1}{(2\pi)^32E_1}\frac{d^3p_3}{(2\pi)^32E_3}
        \frac{d^3p_5}{(2\pi)^32E_5}\frac{d^3p_6}{(2\pi)^32E_6}\nonumber\\
&&\times G^{++}(p_3+p_5+p_6)G^{--}(p_3+p_5+p_6)G^{++}(p_5+p_6)G^{--}(p_5+p_6)
\nonumber\\
&&\times (2\pi)^{4}\delta^{(4)}(p-p_1-p_3-p_5-p_6) \sum_{i=1}^{16} T_i,
\end{eqnarray}
where the $T_i$ are given as
\begin{eqnarray}
T_1 &=& f_q(\vec p) \bar f_q(\vec p_1) \bar f_g(\vec p_3) \bar f_g(\vec p_5) 
        \bar f_g(\vec p_6)\nonumber\\
T_2 &=& f_q(\vec p) \bar f_q(\vec p_1) \bar f_g(\vec p_3) \bar f_g(\vec p_5)
        f_g(-\vec p_6)\nonumber\\
T_3 &=& f_q(\vec p) \bar f_q(\vec p_1) \bar f_g(\vec p_3) f_g(-\vec p_5)
        \bar f_g(\vec p_6)\nonumber\\
T_4 &=& f_q(\vec p) \bar f_q(\vec p_1) \bar f_g(\vec p_3) f_g(-\vec p_5)
        f_g(-\vec p_6)\nonumber\\
T_5 &=& f_q(\vec p) \bar f_q(\vec p_1) f_g(-\vec p_3) \bar f_g(\vec p_5)
        \bar f_g(\vec p_6)\nonumber\\
T_6 &=& f_q(\vec p) \bar f_q(\vec p_1) f_g(-\vec p_3) \bar f_g(\vec p_5)
        f_g(-\vec p_6)\nonumber\\
T_7 &=& f_q(\vec p) \bar f_q(\vec p_1) f_g(-\vec p_3) f_g(-\vec p_5)
        \bar f_g(\vec p_6)\nonumber\\
T_8 &=& f_q(\vec p) \bar f_q(\vec p_1) f_g(-\vec p_3) f_g(-\vec p_5)
        f_g(-\vec p_6)\nonumber\\
T_9 &=& f_q(\vec p) f_{\bar q}(-\vec p_1) \bar f_g(\vec p_3) \bar f_g(\vec p_5)
        \bar f_g(\vec p_6)\nonumber\\
T_{10}&=& f_q(\vec p) f_{\bar q}(-\vec p_1)\bar f_g(\vec p_3)\bar f_g(\vec p_5)
        f_g(-\vec p_6)\nonumber\\
T_{11}&=& f_q(\vec p) f_{\bar q}(-\vec p_1)\bar f_g(\vec p_3)f_g(-\vec p_5)
        \bar f_g(\vec p_6)\nonumber\\
T_{12}&=& f_q(\vec p) f_{\bar q}(-\vec p_1) \bar f_g(\vec p_3) f_g(-\vec p_5)
        f_g(-\vec p_6)\nonumber\\
T_{13}&=& f_q(\vec p) f_{\bar q}(-\vec p_1) f_g(-\vec p_3) \bar f_g(\vec p_5)
        \bar f_g(\vec p_6)\nonumber\\
T_{14}&=& f_q(\vec p) f_{\bar q}(-\vec p_1) f_g(-\vec p_3) \bar f_g(\vec p_5)
        f_g(-\vec p_6)\nonumber\\
T_{15}&=& f_q(\vec p) f_{\bar q}(-\vec p_1) f_g(-\vec p_3) f_g(-\vec p_5)
        \bar f_g(\vec p_6)\nonumber\\
T_{16}&=& f_q(\vec p) f_{\bar q}(-\vec p_1) f_g(-\vec p_3) f_g(-\vec p_5)
        f_g(-\vec p_6).
\end{eqnarray}
By attributing unbarred
functions $f$ to incoming particles and barred functions $\bar f$ to
outgoing ones, one can make the identification of 
$T_{16}$ corresponds with the annihilation process $q\bar q ggg\to ${\O},  
$T_1$ with $q\to qggg$, $T_8$ with $qggg\to q$ while
$T_{12}$, $T_{14}$ and $T_{15}$ correspond to the process
 $q\bar qgg\to g$. These
processes are all kinematically forbidden.
Only processes with at least two partons in both the initial and the final
state are kinematically allowed:
these are  $T_2$, $T_3$ and $T_5$, which lead to $qg\to qgg$, $T_4$, $T_6$
and $T_7$, which lead to $qgg\to qg$, $T_{10}$, 
$T_{11}$ and $T_{12}$, corresponding to $q\bar qg\to
gg$, and finally $T_9$, which gives rise to $q\bar q \to ggg$.

\section{Remaining self energy diagrams}
In Section IIIC, we gave the Feynman graphs which lead to the square of
the individual scattering amplitudes.    There are many diagrams which give
rise to the mixed terms that also are necessary when forming the square
of the full amplitude.   For completeness, we indicate
 which  generic diagrams  also contribute to the self energy
$\Sigma^{(3)}$ and lead to these 
mixed terms between the scattering amplitudes
of the individual channels  $q\bar q \to ggg$, $qg\to qgg$,
$q\bar qg\to gg$ and $qgg\to qg$.   These are given in Fig.~16.
The last six diagrams are not symmetric, and it is to be understood that
the mirror reflected diagrams must also be taken into account.

\end{appendix}

\eject

FIGURES\\
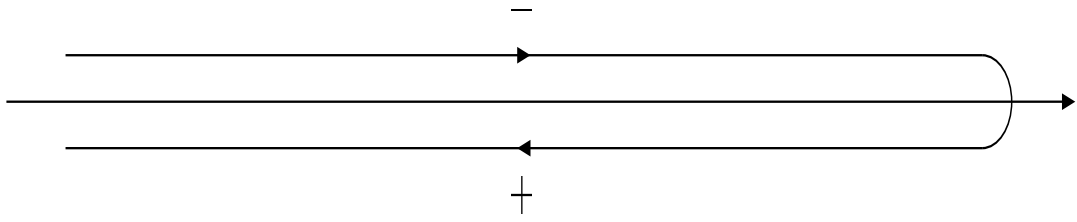
Fig.~1. Closed time path.\hfill \\
Fig.~2.   Quark and gluon generic 
Hartree self energies.  Solid lines
refer to quarks, wavy lines to gluons.\hfill \\
Fig.~3.  Quark and gluon generic one loop self energies.
The quark self energy plus the first gluon self energy are Fock diagrams,
while (b) of this figure is a polarization insertion.\hfill \\
Fig.~4.  Generic diagrams for the quark and gluon self energies
 that contain two loops.\hfill\\
Fig.~5. The rainbow graphs.\hfill\\
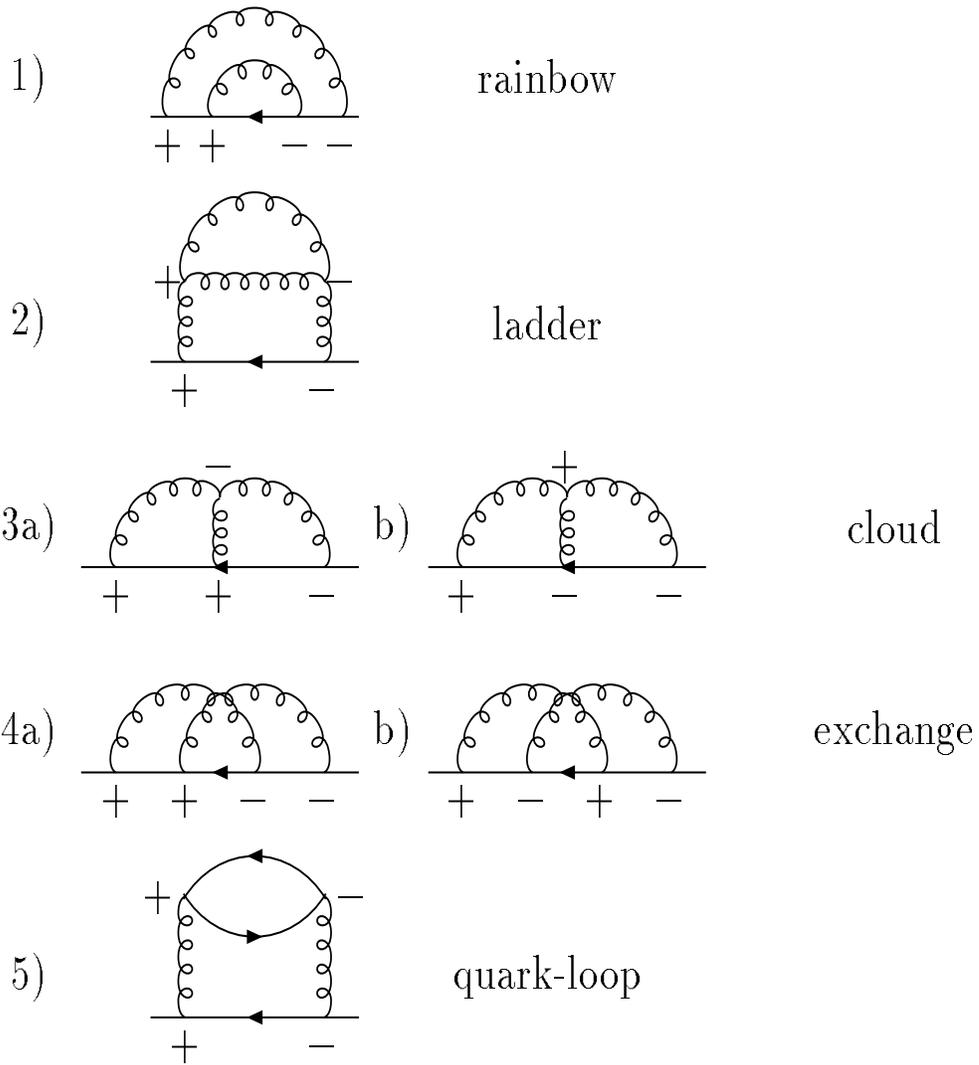
Fig.~6. All diagrams contributing to $\Sigma^{(2)+-}$.\hfill\\
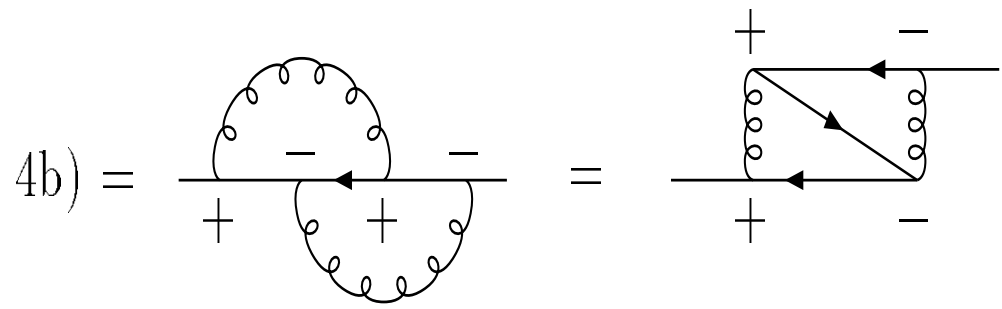
Fig.~7. Diagram of Fig.~6.4b.\hfill\\
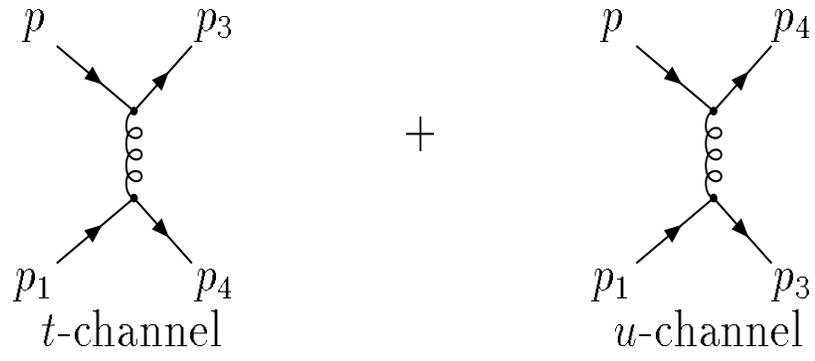
Fig.~8. The $t$ and $u$ channel Feynman graphs for elastic
quark-quark scattering.\hfill\\
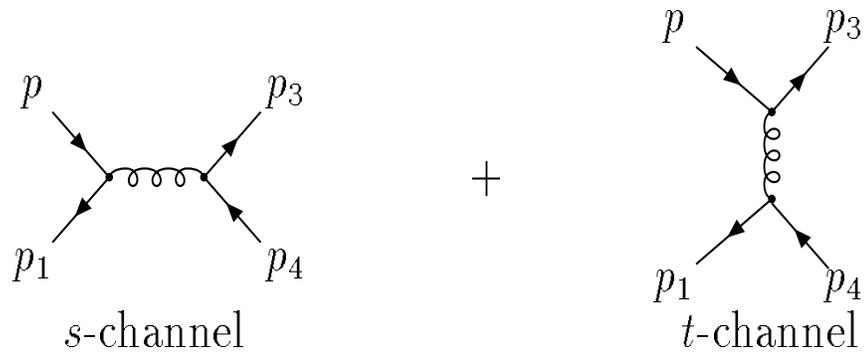
Fig.~9.  The $s$  and $t$ channel Feynman graphs for elastic
quark-antiquark scattering.\hfill\\
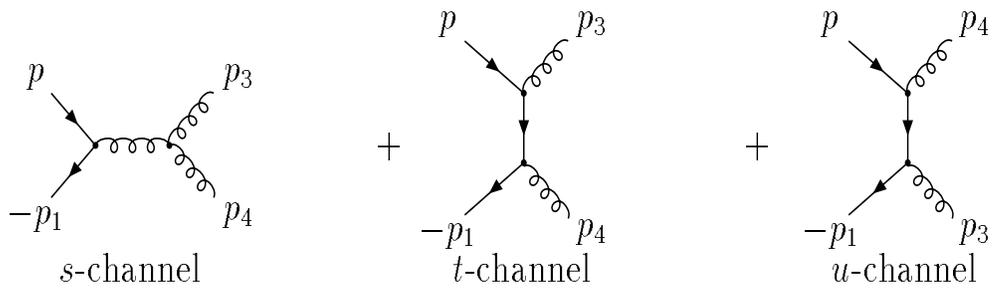
Fig.~10. The $s$, $t$ and $u$ channel Feynman graphs for the
process $q\bar{q}\to gg$.\hfill\\
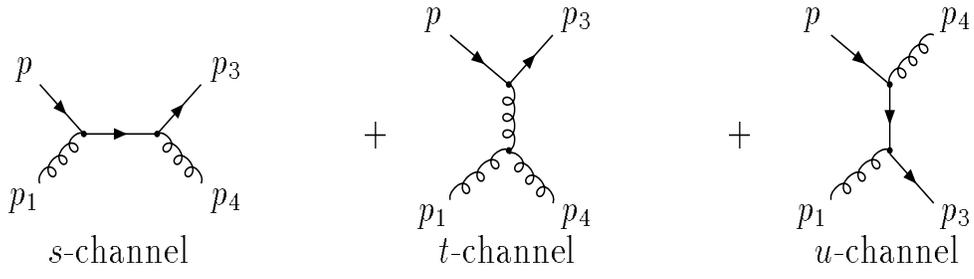
Fig.~11. The $s$, $t$ and $u$ channel Feynman graphs for the
process $qg\to qg$.\hfill\\
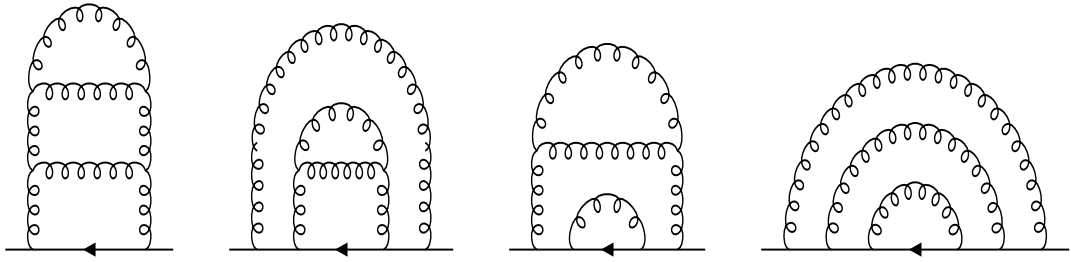
Fig.~12. Generic diagrams for the quark self energy that contain three
exchanged gluons and which give rise to squared scattering amplitudes for
individual channels.\hfill\\
Fig.~13. Feynman graphs for the process $q\bar{q}\to ggg$.\hfill\\ 
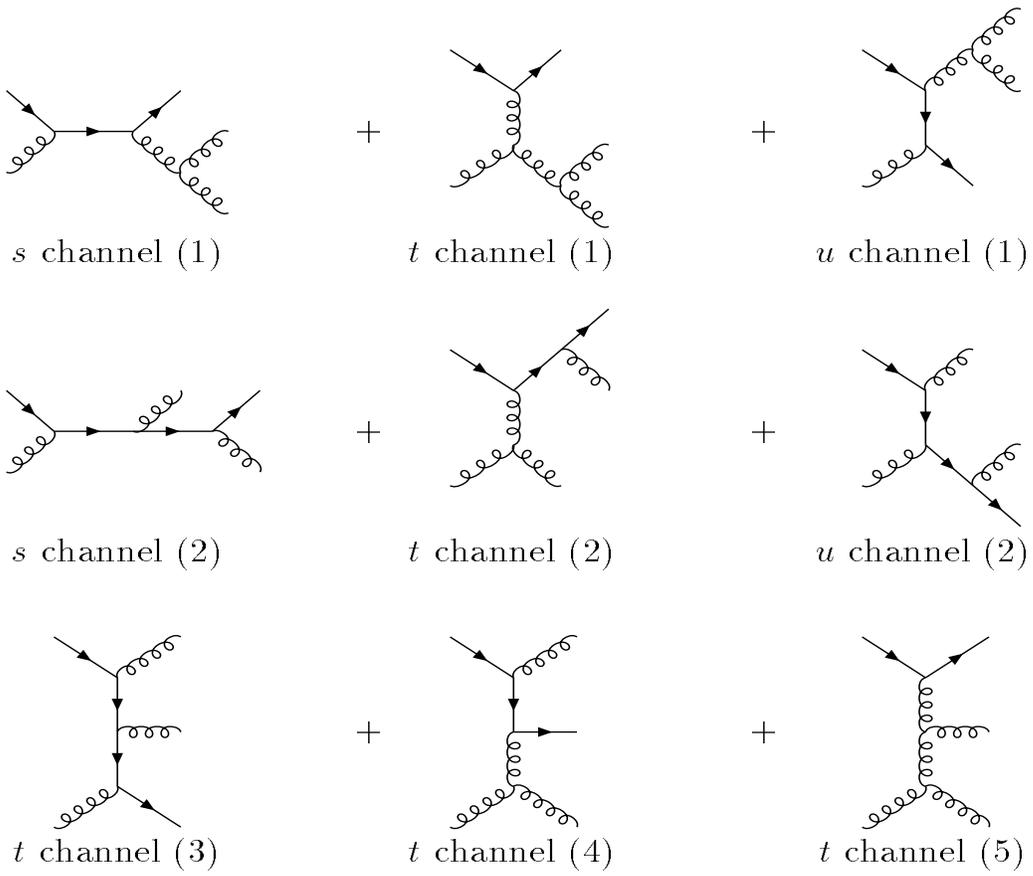
Fig.~14. Feynman graphs for the process $qg\to qgg$.\hfill\\
Fig.~15. The only three rung ladder contribution to the quark self
energy.\hfill\\ 
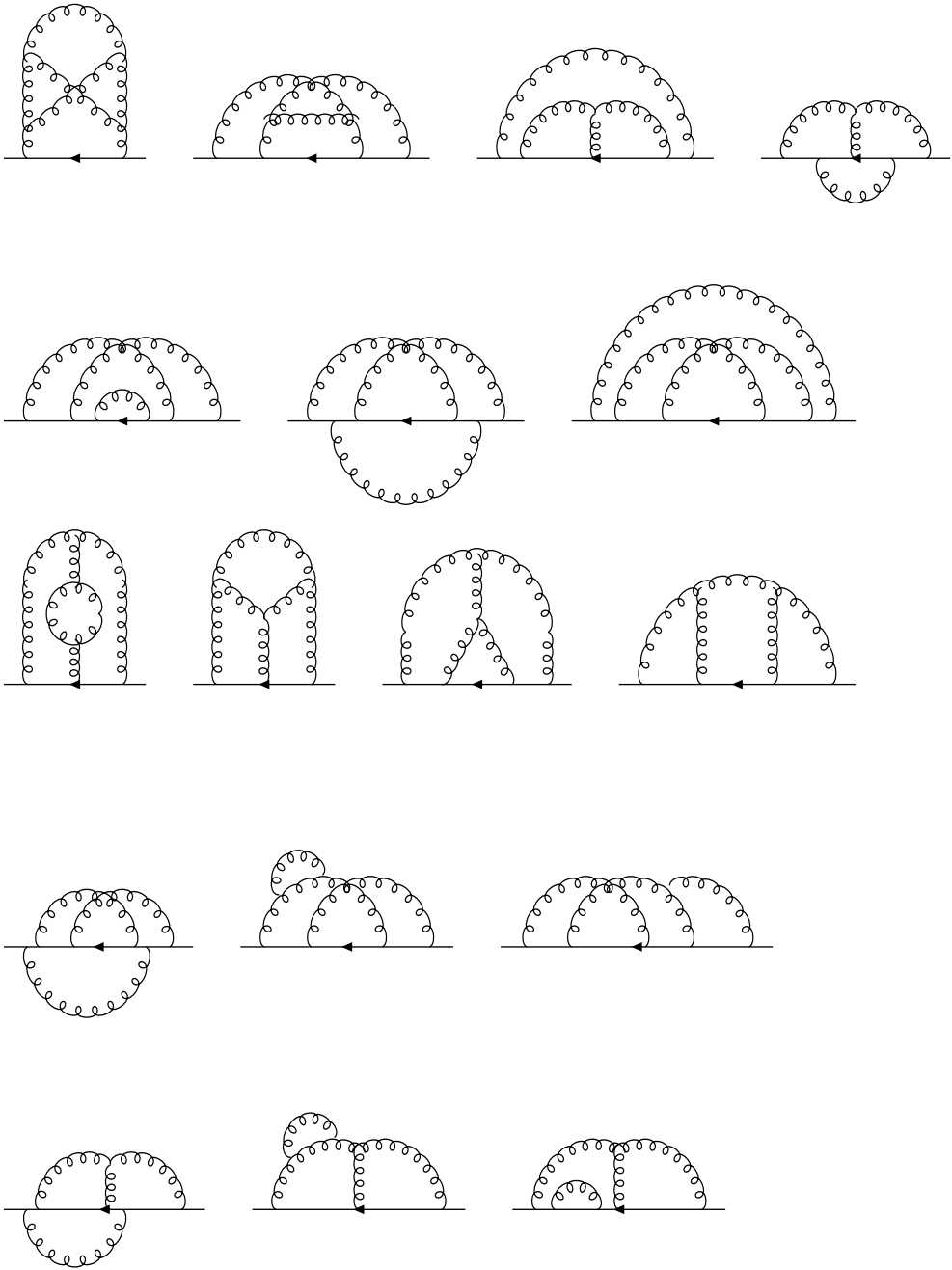
Fig.~16. Generic graphs for the quark self energy 
that are required for constructing the mixed 
amplitudes that lead to scatterings with three gluons.
\end{document}